\documentclass[aps,showpacs,amsmath,amssymb]{revtex4}
\usepackage{graphicx}
\usepackage{bbold}
\usepackage{yfonts}
\usepackage{cancel}
\usepackage{ulem}
\usepackage{amscd}

\usepackage{graphicx}
\providecommand{\color}[2][1]{} 
\usepackage{xcolor} 
\usepackage{color}

\usepackage{dcolumn}
\usepackage{epsfig}
\usepackage{verbatim}

\def\be{\begin{equation}}
\def\ee{\end{equation}}

\newcommand{\ba}{\begin{array}}
\newcommand{\ea}{\end{array}}
\newcommand{\bea}{\begin{eqnarray}}
\newcommand{\eea}{\end{eqnarray}}

\newcommand{\tr}{\mbox{Tr}}                                        
\newcommand{\bra}[1]{\ensuremath{\langle #1 |}}
\newcommand{\ket}[1]{\ensuremath{| #1 \rangle}}

\newcommand{\matel}[3]{\ensuremath{\langle #1 | #2 | #3 \rangle}}
\newcommand{\moy}[1]{\ensuremath{\langle #1 \rangle}}
\newcommand{\op}[1]{\ensuremath{\widehat{ #1} }}
\newcommand{\der}{\partial}
\newcommand{\vct}[1]{\ensuremath\mbox{\boldmath$ #1 $}}
\newcommand{\mat}[1]{\ensuremath\mbox{$ \mathcal #1 $}}

\newcommand{\pim}{{p}}
\newcommand{\qim}{{q}}

\newcommand{\psp}{p^\star}
\newcommand{\qsp}{q^\star}

\newcommand{\pspp}{p^\star_+}
\newcommand{\qspp}{q^\star_+}

\newcommand{\pspm}{p^\star_-}
\newcommand{\qspm}{q^\star_-}

\newcommand{\Qsp}{Q^\star}
\newcommand{\qunsp}{q_1^\star}
\newcommand{\qdeuxsp}{q_2^\star}
\newcommand{\Ssp}{S^\star}
\newcommand{\Sspp}{S^\star_+}
\newcommand{\Sspm}{S^\star_-}

\newcommand{\uc}{\tau}
\newcommand{\dimt}{\ensuremath{\Delta\tau}}

\newcommand{\ti}{t_0}
\newcommand{\tf}{T}

\newcommand{\symb}[1]{\ensuremath\left[ #1 \right]_{\textrm{W}}}
\newcommand{\0}{\textrm{i}}
\newcommand{\1}{\textrm{f}}
\newcommand{\pW}{\mathbb W}
\newcommand{\improp}[2]{\ensuremath {\mathcal C}^{[ #1]}_{#2}}
\newcommand{\reprop}[2]{\ensuremath {\mathcal R}_{#1,#2}}
\newcommand{\map}[3]{\ensuremath {\mathcal M}_{#1,#2,#3}}
\newcommand{\impropA}[1]{\ensuremath {\mathcal C}_{#1}}
\newcommand{\mapA}[1]{\ensuremath {\mathcal M}_{#1}}

\begin{document}

\title{Semi-classical work  and quantum work identities in Weyl representation} 

\author{O.~Brodier$^{1}$, K.~Mallick$^{2}$ and A.~M.~Ozorio de Almeida$^{3}$}
\affiliation{$^1$ Institut Denis Poisson, Campus de Grandmont, Universit\'e de Tours, 37200 TOURS \\
$^2$ Institut de Physique Th\'eorique,,  Universit\'e Paris-Saclay,
CEA and CNRS, 91191 Gif-sur-Yvette, France\\
$^3$ Centro Brasileiro de Pesquisas F\'isicas, 
Rua Dr. Xavier Sigaud, 150, 22290-180 Rio de Janeiro, BRASIL
}

\date{\today}

\begin{abstract}
  We derive a semi-classical nonequilibrium  work  identity  by applying the
  Wigner-Weyl quantization scheme to the Jarzynski identity for a
  classical Hamiltonian. This allows us to extend the concept of work 
to the leading order in $\hbar$. We propose  a geometric interpretation of this 
  semi-classical  Jarzynski relation in terms of trajectories in a complex phase space
  and illustrate it with the exactly solvable case of the quantum harmonic
  oscillator.
\end{abstract}


\maketitle

According to the  second principle of thermodynamics, macroscopic  phenomena tend to evolve towards states corresponding
 to a maximum number of underlying  microstates, i.e. states that maximize entropy. Combined with the first principle, this leads to a more operational statement for  isothermal processes: the  minimal amount of work to modify a system from a state A to a state B is  given by 
\be
W \geq F(B)-F(A),
\ee
where $F=U-TS$ is the free energy. Statistical mechanics has shown that the interpretation of  macroscopic thermodynamics  is statistical, by endowing  microscopic states with a probability measure. Hence,  the second principle should be understood  as  an average of a random process; one should write, in fact,  
\be
\moy W \geq F(B)-F(A),
\ee
where  $\moy W$ is the  average of path-dependent  work along the realizations of a given macroscopic process
 (or protocol) leading from state $A$ to state $B$.
In 1996, C.  Jarzynski \cite{Jar97,JarzPRE} discovered  that
there exists  a non-equilibrium work relation, 
underlying the long-established  work inequality~: 
\be
\moy {e^{-\beta W}} = e^{-\beta (F(B)-F(A))}.
\label{jar1}
\ee
A remarkable consequence, from the thermodynamics point of view, is that, for (\ref{jar1}) to be true, some individual realizations of the transform must `violate' the second principle, that is, the system can reach
once in a while a final state whose free energy variation is actually greater than the received work. The
Jarzynski identity and its generalization by G. Crooks \cite{Crooks99}
have triggered an immense amount of work during the last two  decades
(see  e.g.  \cite{jarzynskiEPJ,JarzReview,SeifertRV}  for reviews) and has been verified
experimentally \cite{Bustamante,ritort,Sergio}.

The original approach of  Jarzynski was  based on classical
trajectories and on the classical definition
of work.  It was  therefore a challenge to  generalize it to a quantum system.
 For  closed  quantum systems,  this difficulty was
overcome by the so-called two measurements
process \cite{kurchan,tasaki,Smukamel,maes,Qjarz,Gaspard1,Hang09,Imry12}, where
work is defined as
the difference of energy between the end and the beginning of the evolution. This scheme was also studied using the Weyl formalism \cite{Liu18,Liu19} and path integrals \cite{Funo18}. As
explained by Talkner et al. \cite{Hang07}, this definition of work 
does  {\it not} correspond to a quantum  observable
because it can not be  represented by a Hermitian operator.  This  explains
why alternative proposals based on some quantum work operator did not  obey
the  Jarzynski identity \cite{yukawa,Nieuw05,Ande16}.  For an open system, the 
two measurements scheme could be applied by considering the system together
with its environment as a global, closed, system (see e.g.  \cite{MukRV,HangRV,Aberg}).
 A different strategy to study open systems is to use a  quantum map that
 acts  on the density matrix of the system  \cite{CrooksQ1,CrooksQ2,Parr15}.
 Under suitable assumptions, this map leads to a quantum Markov evolution,
 described by a Lindblad equation \cite{Haroche}. In this dynamical framework,
 a quantum analog of the  Jarzynski relation  can be proved
 by defining  a work operator through  a generalization of  the Feynman-Kac formula
 to quantum Markov semi-groups  (see \cite{CheMal} and references therein).
 Further studies and proposals  for   experimental
 checks of the  quantum Jarzynski identity have unveiled the  interplay between
 measurement, quantum trajectories  and  stochastic thermodynamics
 \cite{Lutz08,Lesa10,Lutz11,Mahl11,Koslo13,Paz14,Serra15,Hang16,Niss16,Auff16,Vinjana,Rast13, Rast14,Smith18}.

 In  \cite{Rah15}, C. Jarzynski, H. T. Quan and S. Raav studied
 the semi-classical limit of the two measurements process to study the
 correspondence between the quantum and the classical definitions
 of work, and between the corresponding work distributions (see also \cite{Wisniacki17A,Wisniacki17B}).
 The aim of the present work is to revert the logic and to
 derive a quantum Jarzynski identity 
 by using the Weyl representation of quantum mechanics. The advantage of
 this approach is to  restore a classical phase space, allowing us, 
 in  the semi-classical regime, to  define a pseudo-work along
 pseudo-trajectories, whose classical limit  coincides
 with the  definition traditional  work along the classical trajectories.
 Our  semi-classical definition  of the
  work does not require the system to be closed
  and can be associated with a continuously measuring  environment, such
  as modeled by Lindblad type equations, for which nonequilibrium
  work identities are valid \cite{CheMal}.

  The outline of this work is the following. In section~\ref{sec:WWeyl},
  we review the basic properties of the Wigner-Weyl quantization
  scheme that will be used afterwards.
  In section~\ref{sec:semiclass},   we consider  a quantum  system 
  in thermal equilibrium at inverse temperature $\beta$
   governed by a time-independent 
   Hamiltonian:
   starting from  the Wigner transform of the  density  matrix, we
   define  a  'pseudo-Hamiltonian' which,   in the semi-classical limit,
    can be viewed as the average of the true
   classical Hamiltonian over a trajectory in  complex time,  of duration 
   $\Delta\uc =  -i\hbar\beta$.  In section~\ref{WeylVanVleck}, we
   study a system with a time-dependent Hamiltonian: we define a pseudo-work, 
   interpret it  as the time-integral of the power generated by the
   pseudo-Hamiltonian over a complex trajectory and show that this  pseudo-work
   satisfies  semi-classical Jarzynski identity. This relation is illustrated
   by an explicit calculation for the harmonic oscillator
   in section~\ref{harmosc}. The last section is devoted to concluding remarks.

  \section{A brief review of Weyl quantization}
  \label{sec:WWeyl}

In this section, we recall some basic properties of the Wigner-Weyl
quantization scheme \cite{Wigner32,Weyl31}
that we shall use in the present work.
Elementary presentations can be found
in \cite{Case,Curt,Varad} and more advanced discussions  in
\cite{Ber77,Alm98}.

The Weyl transform allows us to construct  an operator
from a phase space function $f(p,q)$ . The idea is simply to take the Fourier transform of $f$ and then to take a modified inverse Fourier transform, where the variables $p$ and $q$ are  replaced by  the operators $\op p$ and $\op q$. Literally, one has
\be
\op f = \frac{1}{\left(2\pi\hbar\right)^2}\iint e^{\frac{i}{\hbar}\left( k_p \op p + k_q \op q\right) }
\left( \iint f(\tilde p,\tilde q) e^{-\frac{i}{\hbar}\left( k_p \tilde p + k_q \tilde q\right) } ~d\tilde p~d\tilde q \right)~dk_p~dk_q.
\ee
The integral gets simpler in the position representation, and, by applying Baker Campbell Hausdorff and the closure relation,
\begin{align}
e^{\frac{i}{\hbar}\left( k_p \op p + k_q \op q\right) } &= e^{\frac{i}{\hbar}\frac{1}{2} k_q \op q  } ~e^{\frac{i}{\hbar} k_p \op p} ~e^{\frac{i}{\hbar}\frac{1}{2} k_q \op q  } \cr
~ &= \int e^{\frac{i}{\hbar}\frac{1}{2} k_q \op q  } ~\ket p e^{\frac{i}{\hbar} k_p p} \bra p ~e^{\frac{i}{\hbar}\frac{1}{2} k_q \op q  } ~dp,
\end{align}
 one ends up with
\be
\matel {q'} {\op f} {q''} = \frac{1}{2\pi\hbar}\int e^{\frac{i}{\hbar}p\left(q'-q''\right)} f\left(p,\frac{q'+q''}{2}\right)~dp.
\label{WQ}
\ee
This relation can be easily inverted, and, from any Hermitian operator $\op A$, one can define its "Weyl symbol" $\symb{\op A}(p,q)$ by
\be
\symb{\op A}(p,q) = \int e^{-\frac{i}{\hbar}pQ} \matel {q+\frac{Q}{2}}{\op A}{q-\frac{Q}{2}}~dQ,
\label{weyltrans}
\ee
which is a function of classical phase space. The Weyl symbol of a Hermitian operator is real, as can be seen easily by taking the complex conjugate of (\ref{weyltrans}). This integral is sometimes called the Wigner transform. When the operator is a density operator, one adds a prefactor for normalization
\be
W(p,q) = \symb{\op A}(p,q) = \frac{1}{2\pi\hbar}\int e^{-\frac{i}{\hbar}pQ} \matel {q+\frac{Q}{2}}{\op \rho}{q-\frac{Q}{2}}~dQ,
\label{wigner}
\ee

Since the transform is one-to-one, the Weyl representation is strictly equivalent to regular quantum mechanics. For  operators made of a single variable, it actually respects the "correspondence principle". For instance one has
\begin{align}
\symb{\op p^n}(p,q)  &= p^n \cr
\symb{\op q^n}(p,q)  &= q^n.
\end{align}
On the other hand, the Weyl symbol of a product of non-commuting operators is generally not the product of the Weyl symbols of the operators. One has in fact
\be
\symb{\op A\op B}(p,q) = \frac{1}{\pi^2\hbar^2}\int e^{\frac{2i}{\hbar}\left[(p_1-p)({q_2}-q)- ({p_2}-p)(q_1-q)\right]}A(p_1, q_1)B(p_2,q_2)~dp_1~dq_1~dp_2~dq_2.
\label{productrule2}
\ee
For instance one has
\begin{align}
\left[\op p\op B\right](p,q)&=  \left( p +\frac{\hbar}{2i}\frac{\der}{\der q}\right) B(p,q) \cr
\left[\op q\op p\op B\right](p,q)&=
 \left( q -\frac{\hbar}{2i}\frac{\der}{\der p}\right)\left( p +\frac{\hbar}{2i}\frac{\der}{\der q}\right) B(p,q).
\end{align}
More generally, one has  \cite{Curt} 
\be
\left[\op A\op B\right](p,q) = A\left(  p+\frac{\hbar}{2i}\frac{\der}{\der q},q -\frac{\hbar}{2i}\frac{\der}{\der p}\right) B(p,q).
\label{prodrule1app}
\ee

Although this product rule implies that products of operators are represented by complicated expressions, this simplifies in the case of symmetrized products of operators. For instance one has
\begin{align}
\symb{\op p\op q}(p,q) &= pq + \frac{\hbar}{2i} \cr
\symb{\op q\op p}(p,q) &= pq - \frac{\hbar}{2i},
\end{align}
and consequently,
\be
\symb{\frac{\op p\op q + \op q\op p}{2}}(p,q) = pq.
\ee
This 
problem of  ordering products  disappears as soon as  one takes the trace of a product of operators. Indeed, for every couple of operators $\op A$ and $\op B$ and their corresponding Weyl symbols $A(p,q)$ and $B(p,q)$, one has the following
fundamental identity:
\be
\tr{\op A\op B} = \iint A(p,q) B(p,q)~dp~dq.
\label{prodscalW}
\ee

We emphasize that the Weyl symbol of a general function of a combination of non-commuting operators is generally not the function of the corresponding Weyl symbol:
\begin{align}
\symb{ \exp{\left(\op A\right)} }(p,q) \neq \exp{\left(\symb{\op A}(p,q)\right)},
\label{expweyl}
\end{align}
(unless  the operator $A$  depends on a single variable, that is, $\op A = f(\op p)$ or $\op A = f(\op q)$). In particular, this  implies that
one can not  obtain a Jarzynski equality in the Weyl representation by
simply quantizing the classical Jarzynski proof, which is based on the properties of the exponential function. One of the motivations of the present
work  is to  overcome this difficulty (see  in section \ref{WeylVanVleck}).

\section{Semi-classical approximation of a thermal state}
\label{sec:semiclass}

In  this section, we construct in the classical phase space
a `pseudo-Hamiltonian' $\Gamma(p,q)$,
defined from the Weyl symbol of the thermal state generated by the
quantum Hamiltonian $\op H$. The function   $\Gamma(p,q)$
 is defined in the following way:
\be
\left[ e^{-\beta\op H}\right]_{\textrm{W}}(p,q) \equiv e^{-\beta \Gamma(p,q)}.
\label{defGamma}
\ee
we emphasize again  that, because of  (\ref{expweyl}),
we have $\Gamma(p,q) \neq H(p,q)$, where 
  $H(p,q)$ is the classical Hamiltonian.
The  exact formula  $\Gamma$ is rather complicated but we can derive
an approximate expression $G(p,q) \simeq \Gamma(p,q)$ in the
semi-classical limit, by interpreting the thermal state as a Schr\"odinger
propagator during an imaginary time \cite{Henri,Cartierbook}
\be
e^{-\beta \op H} = e^{-\frac{i}{\hbar}\Delta\uc \op H},
\ee
where
\be
\Delta\uc= -i\hbar\beta
\label{def:imaginet}
\ee
is interpreted as an imaginary time.
For a real $\Delta\uc$, this propagator can be well approximated by
the Van Vleck propagator \cite{chaosbook}, which  plugged into (\ref{wigner}), gives the semi-classical Wigner propagator calculated by M. V. Berry  (see
equation (21) of \cite{berry89}). Therefore, using Berry's result, 
the semi-classical Wigner thermal state is simply
the continuation  of this propagator for imaginary $\Delta\uc$.

In this section, we  rederive the  expression of the  semi-classical Wigner thermal state  by  the  stationary phase method. This will allow us
to introduce   notations  and techniques that will be useful in the rest
of this work. Starting  from equation (38.30) of \cite{chaosbook},
we write 
\be
\matel {q_\1} {e^{-\frac{i}{\hbar}(t_\1-t_\0)\op H}}{q_\0} \simeq K_{\textrm{sc}}(q_\0,t_\0;q_\1,t_\1) \equiv \sum_j \frac{1}{(2i\pi\hbar)^{1/2}}\left| \frac{\der p_\1}{\der q_\0}\right|^{1/2}
e^{\frac{i}{\hbar}S_j(q_\0,t_\0;q_\1,t_\1)-im_j\frac{\pi}{2}},
\label{VanVleck}
\ee
where $S_j$ is a solution of a Hamilton-Jacobi equation or its time reverse,
\be
\frac{\der S_j}{\der t_f} + H\left(\frac{\der S_j}{\der q_f},q_f\right) = 0
\qquad
\frac{\der S_j}{\der t_i} - H\left(-\frac{\der S_j}{\der q_i},q_i\right) = 0,
\label{hamjac}
\ee
and coincides with the classical action calculated along one of the $j^{\textrm{th}}$ classical trajectories $\left(p_j(t),q_j(t)\right)$ generated by $H(p,q)$, the classical counterpart of $\op H$, such that 
\be
\left\{
\begin{aligned}
q_j(t_\0) &= q_\0 \cr
q_j(t_\1) &= q_\1 
\end{aligned}
\right.
\qquad
\left\{
\begin{aligned}
\der_{t}p_j(t) &= -\der_qH\left(p_j(t),q_j(t)\right) \cr
\der_{t}q_j(t) &= \der_pH\left(p_j(t),q_j(t)\right),
\end{aligned}
\right.
\label{dotx}
\ee
There are a priori several classical trajectories, labelled by $j$, and the propagator (\ref{VanVleck}) sums over all the trajectories having the right initial and final conditions. The integers  $m_j$ in equation~(\ref{VanVleck})
are the Maslov indices, which count the number of times the $j^{th}$
trajectory crosses a  {\it  caustic}, that is, the set of points where $\frac{\der p_\1}{\der q_\0}$ is singular \cite{chaosbook,Maslov}. These caustics are reached only after a certain amount of time;  before that time, there is
a  unique trajectory for two given boundary conditions, here $(q_\0,t_\0)$ and $(q_\1,t_\1)$. Thus, we have 
\begin{align}
S(q_\0,t_\0;q_\1,t_\1) &= \int_{t_\0}^{t_\1} \left[ p(t)\der_{t}q(t) - H\left(p(t),q(t)\right)\right]~dt \cr
~ &= \int_{t_\0}^{t_\1} p(t)\der_{t}q(t)~dt - (t_\1-t_\0)H\left(p(t_\0),q(t_\0)\right),
\end{align}
the last line being true only for a time independent Hamiltonian.
We also have, from the fact that $S$ is solution of (\ref{hamjac}), and see chapter 46 of \cite{Arnold} for the whole story,
\be
\left\{
\begin{aligned}
\frac{\der S}{\der q_\0} &= - p_\0 \cr
\frac{\der S}{\der q_\1} &= p_\1 \label{pipf}
\end{aligned}
\right.
\left\{
\begin{aligned}
\frac{\der S}{\der t_\0} &= H(p_\0,q_\0) \cr
\frac{\der S}{\der t_\1} &= -H(p_\1,q_\1) \, .
\end{aligned}
\right.
\ee
We now  take the analytical continuation of this propagator and obtain 
\begin{align}
\matel {q_\1}{e^{-\beta \op H}}{q_\0} &\simeq K_{\textrm{sc}}(q_\0,\uc_\0;q_\1,\uc_\1) \cr
~ &\simeq \frac{1}{(2i\pi\hbar)^{1/2}}\left| \frac{\der^2 S}{\der q_\1\der q_\0}\right|^{1/2}
e^{\frac{i}{\hbar}S(q_\0,\uc_\0;q_\1,\uc_\1)} \, ,
\label{semictherm}
\end{align}
with $\uc_\0 \in i\mathbb R$ and $\uc_\1=\uc_\0+\dimt\in i\mathbb R$ purely imaginary times.
We have  supposed that  all  Maslov indices $m_j$ vanish: this 
 requires  the imaginary time to be small enough to avoid the first caustics of the complex trajectory. Like in regular WKB method, the caustics corresponds to the singularities of the second derivative of the Jacobian of $S$, and the Maslov indices increase by one every time the trajectory crosses a caustic. It is a consequence of the necessary change of represensation, from $(p,q)$ to the Fourier conjugate variables $(\xi_p,\xi_q)$, in order to have a non singular expression of the propagator in the neighbourhood of the caustic. This procedure is more delicate in the complex domain than in the real one, as one needs to chose the "correct branch" connecting the trajectory from both sides of the caustic. This is the Stokes phenomenon.

To obtain a correct ordering of the $\hbar$ corrections, the value of $\Delta\uc=\uc_\1-\uc_\0$ must 
remain  constant as $\hbar\rightarrow 0$ \footnote{This will  ensure that the prefactor $\frac{1}{(2i\pi\hbar)^{1/2}}\left| \frac{\der^2 S}{\der q_\1\der q_\0}\right|^{1/2}$ is  of higher order in $\hbar$ than the  exponential of $S$.}. This corresponds to a semiclassical regime at low temperature ( large $\beta$ ). Although the limit itself, $\hbar=0$, cannot really be interpreted as a classical regime, since it must occur at $0$ temperature which is the realm of quantum regime, we can expect that finite values of $\hbar$ and finite values of temperature can be in this intermediate regime. 
We remark  that the initial imaginary "time" $\uc_\0\in i\mathbb R$ is, {\it a priori}, a free parameter, since the Hamiltonian is not a function of $\beta$. Let us now  have a closer look at the action $S$: we consider 
 the imaginary time classical trajectory $\left(\pim(\uc),\qim(\uc)\right)$ generated by $H(p,q)$, such that
\be
\left\{
\begin{aligned}
\qim(\uc_\0) &= q_\0 \cr
\qim(\uc_\1) &= q_\1
\end{aligned}
\right.
\qquad
\left\{
\begin{aligned}
\der_{\uc}\pim(\uc) &= -\der_qH\left(\pim(\uc),\qim(\uc)\right) \cr
\der_{\uc}\qim(\uc) &= \der_pH\left(\pim(\uc),\qim(\uc)\right),
\label{condtraj}
\end{aligned}
\right.
\ee
where $\uc$ is an imaginary parameter with $\uc\in [\uc_\0,\uc_\1]$. This trajectory is generically unique and 
the action $S$ (which is now imaginary)
is  calculated by integrating  along  this  trajectory (\ref{condtraj}),
\begin{align}
S(q_\0,\uc_\0;q_\1,\uc_\1)  &= \int_{\uc_\0}^{\uc_\1} \left[ \pim(\uc)\der_{\uc} {\qim}(\uc) - H\left(\pim(\uc),\qim(\uc)\right)\right]~d\tau \cr
~ &= \int_{\uc_\0}^{\uc_\1} \pim(\uc)\der_{\uc} {\qim}(\uc) ~d\tau - \dimt H\left(\pim(\uc_\0),\qim(\uc_\0)\right).
\label{thermaction}
\end{align}

We now evaluate (\ref{defGamma}) by plugging  the semi-classical expression
(\ref{semictherm}) in the Wigner transform (\ref{wigner}):
\begin{align}
\left[e^{-\beta \op H}\right]_{\textrm {W}}(p,q) &= \frac{1}{2\pi\hbar}\int e^{-\frac{i}{\hbar}pQ}\matel {q+\frac{Q}{2}} {e^{-\beta \op H_t}} {q-\frac{Q}{2}}~dQ \cr
~ &\simeq \frac{1}{2\pi\hbar}\int \frac{1}{(2i\pi\hbar)^{1/2}}\left| \frac{\der^2 S}{\der q_\1\der q_\0}\right|^{1/2} e^{\frac{i}{\hbar}\left[ S_{\textrm{tot}}(p,q,Q) \right]}~dQ \, .
\label{semicthermal2}
\end{align}
 For every $Q$,   the total action
\be
S_{\textrm{tot}}(p,q,Q) = -pQ + S(q-\frac{Q}{2},\uc_\0;q+\frac{Q}{2};\uc_\1),
\label{tot:action}
\ee
defines, implicitly, a classical trajectory $(p_\0,q_\0) \rightarrow (p_\1,q_\1)$ such that 
\begin{align}
q_\0 &= \qim\left(\uc_\0\right) = q-\frac{Q}{2} \cr
q_\1 &= \qim\left(\uc_\1\right) = q+\frac{Q}{2}.
\label{qiqf}
\end{align}
In the semi-classical limit, $\hbar\rightarrow 0$, keeping $\dimt$ fixed,
the  stationary phase method can be used;  the main contribution in
the integral  (\ref{semicthermal2})  is
given by  the stationary point $\Qsp$, that  solves 
\begin{align}
\frac{\der}{\der Q} S_{\textrm{tot}}(p,q,Q) \Huge|_{\Qsp} =
- \frac{1}{2}\frac{\der S}{\der q_\0}\left(q-\frac{\Qsp}{2},\uc_\0;q+\frac{\Qsp}{2};\uc_\1\right)
+ \frac{1}{2}\frac{\der S}{\der q_\1}\left(q-\frac{\Qsp}{2},\uc_\0;q+\frac{\Qsp}{2};\uc_\1\right) - p &= 0.
\label{spp1}
\end{align}
Thus, according to (\ref{pipf}), we obtain 
\be
p = \frac{p_\0+p_\1}{2}.
\label{pmid}
\ee
Using equations (\ref{qiqf}) and (\ref{pmid}),
 we  then conclude that 
 among the family of trajectories $(\pim(\uc),\qim(\uc))$ spanned by $Q$,
 we must select the stationary  trajectory $(\psp(\uc),\qsp(\uc))$ such that
\begin{align}
\frac{\psp_\0 + \psp_\1}{2} &= p \label{midpoint-p} \\
\frac{\qsp_\0 + \qsp_\1}{2} &= q, \label{midpoint-q}
\end{align}
where $(\psp_\0,\qsp_\0) = (\psp(\uc_\0),\qsp(\uc_\0))$ and $(\psp_\1,\qsp_\1) = (\psp(\uc_\1),\qsp(\uc_\1))$.
To understand the structure of the solution, let us define the imaginary time flow $(p',q')\mapsto (p(\uc),q(\uc))=\impropA {\uc} (p',q')$ with $(p(\uc),q(\uc))$ solution of equation (\ref{condtraj}) and $(p(0),q(0))=(p',q')$. Formally, one can write
\begin{align}
p(\uc)&=p(0) + \uc \dot p(0) + \frac{1}{2}\uc^2 \ddot p(0) + \ldots \cr
q(\uc)&=q(0) + \uc \dot q(0) + \frac{1}{2}\uc^2 \ddot q(0) + \ldots 
\end{align}
We first note  that, for real $(p(0),q(0))$, then $\dot p(0)$, $\ddot p(0)$ \ldots are also real, as they can be obtained from (\ref{dotx}) and expressed in terms of derivatives of the type $\der_{p^nq^m}H(p(0),q(0))$ where $H$ is real. On the other hand, $\uc$ is imaginary. Therefore one then has, for real $(p',q')$,
\be
\overline{\impropA {\uc}(p',q')} = \impropA {-\uc}(p',q').
\label{conjmap}
\ee
Let us then build the map
\be
(p',q')\mapsto {\mapA{\frac{\dimt}{2}}}(p',q')=\frac{\impropA{\frac{\dimt}{2}}(p',q')+\impropA{-\frac{\dimt}{2}}(p',q')}{2}.
\ee
From (\ref{conjmap}), ${\mapA{\frac{\dimt}{2}}}$ is a real map from the real phase space $\left\{(p',q')\right\}$ to itself. Then, we  define
\be
(\psp_c,\qsp_c) = \left({\mapA{\frac{\dimt}{2}}}\right)^{-1}(p,q).
\ee
Obviously,
\begin{align}
(\psp_\0,\qsp_\0) &= \impropA{-\frac{\dimt}{2}}(\psp_c,\qsp_c) \cr
(\psp_\1,\qsp_\1) &= \impropA{\frac{\dimt}{2}}(\psp_c,\qsp_c)
\end{align}
then fulfill conditions (\ref{midpoint-p}) and (\ref{midpoint-q}).
This construction makes it clear that
\begin{align}
\psp_\0 &= \overline{\psp_\1} \cr
\qsp_\0 &= \overline{\qsp_\1},
\end{align}
which implies, from equation (\ref{qiqf}) with $q$ being real,
that $\frac{\Qsp}{2}$ is in fact  the imaginary part of $q_\1$, with
\be
\overline{\Qsp} = -\Qsp.
\ee

\begin{figure}
\begin{center}
\scalebox{0.5}{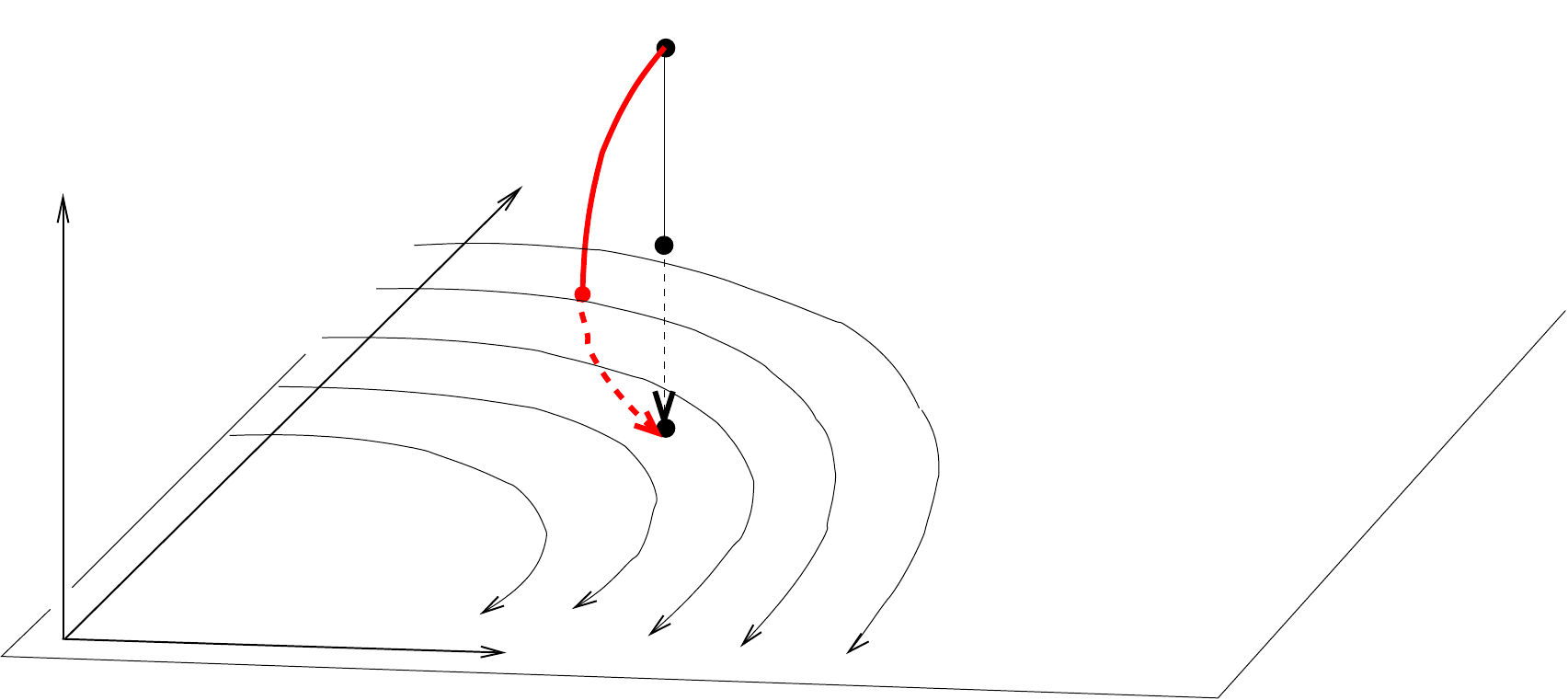}
\caption{The complex trajectory crosses the real plane at $(\psp_c,\qsp_c)$; its chord $(\psp_\1-\psp_\0,\Qsp)$ is imaginary and the middle of this chord is $(p,q)$. Also, $(p,q)$ is the image of $(\psp_c,\qsp_c)$ through 
$\mapA {\frac{\dimt}{2}}$.}
\label{sptraj1}
\end{center}
\end{figure}

To summarize, the arc $(\psp(\uc),\qsp(\uc))$ is symmetric with regard to the real phase space plane, and it intersects this real phase space plane at $(\psp_c,\qsp_c)=(\psp(\frac{\uc_\0+\uc_\1}{2}),\qsp(\frac{\uc_\0+\uc_\1}{2}))$. Moreover, the chord $(\psp_\1-\psp_\0,\Qsp)$ of this arc is purely imaginary, and the middle of this chord is $(p,q)$. The picture is shown on figure \ref{sptraj1}.
Finally,  we retrieve
for   the Weyl symbol, defined  in (\ref{semicthermal2}), an 
expression equivalent to the one
in  \cite{berry89}, but  in {\it imaginary time}, that is
\begin{align}
\left[e^{-\beta \op H}\right]_{\textrm {W}}(p,q) &\equiv e^{-\beta \Gamma(p,q)} \mathop{\simeq}_{\hbar\rightarrow 0} \mathcal N(p,q) e^{-\beta G(p,q)} 
\end{align}
with
\begin{align}
G(p,q) &= -\frac{1}{\Delta\uc}\left[
-p\Qsp + S(q-\frac{\Qsp}{2},\uc_\0;q+\frac{\Qsp}{2},\uc_\1)
\right] \cr
~ &= H\left(\psp_c,\qsp_c\right) - \frac{1}{\Delta\uc}\left[  \int_{\uc_\0}^{\uc_\1} 
\psp(\uc)\der_{\uc} \qsp(\uc)  ~d\tau - p\Qsp \right] \cr
~ &= H\left(\psp_c,\qsp_c\right) - \frac{1}{|\Delta\uc|}A(p,q) \label{expG}
\end{align}
 and the prefactor
\begin{align}
\mathcal N(p,q) &=\frac{1}{2\pi\hbar}\frac{1}{(2i\pi\hbar)^{1/2}}\left| \frac{\der^2 \Ssp}{\der q_\1\der q_\0}\right|^{1/2}\sqrt{\frac{\pi}{-\frac{i}{\hbar}S_{\textrm{tot}}''(\Qsp)}} \cr
~ &=\frac{1}{2\pi\hbar}\frac{1}{(2i\pi\hbar)^{1/2}}\left| \frac{\der^2 \Ssp}{\der q_\1\der q_\0}\right|^{1/2}e^{i\frac{\pi}{4}}
\sqrt{\frac{8\pi\hbar}{\left|\frac{\der^2 \Ssp}{\der q_\1^2}+\frac{\der^2 \Ssp}{\der q_\0^2}-2\frac{\der^2 \Ssp}{\der q_\1\der q_\0}\right|}} \cr
~ &= 
\frac{1}{2\pi\hbar}\sqrt{\frac{2\left|\frac{\der^2 \Ssp}{\der q_\1\der q_\0}\right|}{\left|\frac{\der^2 \Ssp}{\der q_\1\der q_\0}-\frac{1}{2}\left(\frac{\der^2 \Ssp}{\der q_\1^2}+\frac{\der^2 \Ssp}{\der q_\0^2}\right)\right|}} \, .\label{pref}
\end{align}
Here, $\Ssp$  represents the action $S$ evaluated at the saddle-point.
The term  $A(p,q)=i \int_{\uc_\0}^{\uc_\1} \psp(\uc)\der_{\uc} {\qsp}(\uc)  ~d\tau -ip\Qsp$ is a real number that
can be interpreted as the area between the complex arc $(\psp(\uc),\qsp(\uc))$ and its chord, as shown in figure \ref{sptraj1}. As a consequence, $G(p,q)$ is also a real function. On the other hand, $\Gamma(p,q)$ is also real since the Weyl representation of any Hermitian operator is real, as can easily be checked from (\ref{weyltrans}). The prefactor (\ref{pref}) is the product of two terms:  one arises from 
the Van Vleck propagator and the other is generated
by the stationary phase method (i. e. an  imaginary Gaussian integration).
In the following,  we shall keep  the leading order in $\hbar$ only 
and the  prefactor  $\mathcal N(p,q)$ will be omitted.

\section{A  Jarzynski identity in the Weyl representation}
\label{WeylVanVleck}

In the previous section, using a semi-classical approach, 
we have derived  an expression for  the function $G(p,q)$ such
that ${\rm e}^{-\beta G(p,q)}$ represents
the  quantum thermal state generated by the (quantum)
Hamiltonian $\hat H$. We shall now introduce an explicit time dependence
in  the Hamiltonian  $H_t$,  define a pseudo-work,  calculate its 
 semi-classical expression  $\der_t G_t$ and derive a 
 formal Jarzynski identity. 
 The heart of the matter resides in the geometric interpretation of the
 semi-classical trajectories. 

\subsection{Time-dependent Hamiltonian}
\label{timedep}

We  use the semi-classical scheme constructed in  
section~\ref{sec:semiclass},
but with a time dependent Hamiltonian $H_t(p,q)$. In this context,
it is important to be aware that the 'imaginary time' $\uc$ of the trajectory (\ref{condtraj}), which is related to  the temperature $1/\beta$, has nothing to do with the physical time $t$. In particular, the physical time $t$
must remain {\bf frozen during the imaginary time propagation} in (\ref{condtraj}). In other
 words, the imaginary time trajectory $\left(\pim_t(\uc),\qim_t(\uc)\right)$ obeys
\be
\left\{
\begin{aligned}
\qim(\uc_\0) &= q_\0 \cr
\qim(\uc_\1) &= q_\1
\end{aligned}
\right.
\qquad
\left\{
\begin{aligned}
\der_{\uc}\pim(\uc) &= -\der_qH_t\left(\pim(\uc),\qim(\uc)\right) \cr
\der_{\uc}\qim(\uc) &= \der_pH_t\left(\pim(\uc),\qim(\uc)\right),
\end{aligned}
\right.
\label{condtrajt}
\ee
with a fixed value of $t$. This also means that  $\Ssp$ and $\Qsp$ and $\left(\psp(\uc),\qsp(\uc)\right)$ are then functions of $t$.
\vskip 0.3cm
\noindent 
Remark: We emphasize that the trajectory (\ref{condtrajt})
 {\bf is not the analytical continuation of the real trajectory generated by $H_t(p,q)$}. Indeed, such a continuation $\left(\bar p(\uc),\bar q(\uc)\right)$ would obey a  slightly different equation
\be
\left\{
\begin{aligned}
\bar q(\uc_\0) &= q_\0 \cr
\bar q(\uc_\1) &= q_\1
\end{aligned}
\right.
\qquad
\left\{
\begin{aligned}
\der_{\uc}\bar q(\uc) &= -\der_{\bar q}H_\uc\left(\bar p(\uc),\bar q(\uc)\right) \cr
\der_{\uc}\bar q(\uc) &= \der_{\bar p}H_\uc\left(\bar p(\uc),\bar q(\uc)\right).
\label{analytHam}
\end{aligned}
\right.
\ee
Here,  the Hamiltonian  is changing  along the trajectory. 

For any given value of $t$, we  calculate $G_t(p,q)$, the Van Vleck approximation of the pseudo-Hamiltonian $\log{\left(\symb{e^{-\beta \op H_t}}(p,q)\right)}$, by using  equation (\ref{expG}).

We  now  define the {\it  pseudo-work} as $\der_t G_t(p,q)$. The expression of this time derivative is actually simpler
than the expression of $G_t(p,q)$ itself, as we shall now show.
We first  consider  the time-dependent stationary phase trajectory, $(\psp_t(\uc),\qsp_t(\uc))$.
Using equation (\ref{tot:action}), we write 
\be
-\Delta\uc G_t(p,q)=S_{\textrm{tot}}(p,q,\Qsp,t) = -p\Qsp + \Ssp_t\left(q-\frac{\Qsp}{2},\uc_\0;q+\frac{\Qsp}{2},\uc_\1\right),
\ee
where $\Qsp$ is a function of time $t$. Taking derivative with respect to  time,  we obtain
after using (\ref{spp1}):
\begin{align}
-\Delta\uc \der_t G_t(p,q) &= \frac{\der}{\der t}S_{\textrm{tot}}(p,q,\Qsp,t) + \frac{\der}{\der \Qsp}S_{\textrm{tot}}(p,q,\Qsp,t)
\frac{\der\Qsp}{\der t} \cr
~ &= \frac{\der}{\der t}S_{\textrm{tot}}(p,q,\Qsp,t)\cr 
~ &= \frac{\der}{\der t}\Ssp_t\left(q-\frac{\Qsp}{2},\uc_\0;q+\frac{\Qsp}{2},\uc_\1\right) \cr
~ &=\lim_{dt\rightarrow 0} \frac{\Ssp_{t+dt}\left(q-\frac{\Qsp}{2},\uc_\0;q+\frac{\Qsp}{2},\uc_\1\right)-\Ssp_t\left(q-\frac{\Qsp}{2},\uc_\0;q+\frac{\Qsp}{2},\uc_\1\right)}{dt} \, .
\end{align}
We made explicit the latter time derivative so that the reader can remind that $\Ssp_{t+dt}$ and $\Ssp_t$ actually live on two different stationary phase trajectories, with an implicit $t$ dependence. However, the trajectory for time $t+dt$ can be seen as a fluctuation $\left(\pim(\uc)+\delta\pim(\uc),\qim(\uc)+\delta\qim(\uc)\right)$ around the trajectory $\left(\pim(\uc),\qim(\uc)\right)$ for time $t$, therefore,
\begin{align}
-\Delta\uc \der_t G_t(p,q) &= \lim_{dt\rightarrow 0} \frac{1}{dt} \int_{\uc_\0}^{\uc_\1} 
\left[ \left(\pim(\uc)+\delta\pim(\uc)\right)\der_\uc\left(\qim(\uc)+\delta\qim(\uc)\right) - H_{t+dt}\left(\pim(\uc)+\delta\pim(\uc),\qim(\uc)+\delta\qim(\uc)\right) \right. \cr 
~ &~ \left. ~~~~~~~~~~~~~~~~~~ - \pim(\uc)\der_\uc\qim(\uc) + H_{t}\left(\pim(\uc),\qim(\uc)\right) \right]d\uc\, ,
\label{fluctaction}
\end{align}
and, because of the stationarity of the action around $\left(\pim(\uc),\qim(\uc)\right)$ with the same initial and final positions and times, the first order terms in $(\delta \pim(\uc),\delta\qim(\uc))$ cancels out, and only remains the derivative with respect to the explicit time dependence of $H_t$, therefore
\be
\boxed{
\der_t G_t(p,q)
= \frac{1}{\dimt}\int_{\uc_\0}^{\uc_\1} \der_t H_t\left(\psp_t(\uc),\qsp_t(\uc)\right) ~d\tau.
}
\label{derG}
\ee
To summarize,  we have shown that in the semi-classical limit $e^{-\beta \op H_t}$ can be approximately represented by a function $e^{-\beta G_t(p,q)}$ in the Weyl space and that the associated work 
 $\der_t G_t(p,q)$ is given by a simple expression 
as an average of $\der_t H_t(p,q)$ over a complex trajectory.

We note  that
\be
\lim_{\hbar\rightarrow 0} G_t(p,q) \neq H_t(p,q),
\ee
because, when $\dimt$ is fixed, the $\hbar\rightarrow 0$ limit
has to be taken together with $\beta\rightarrow +\infty$: this  is not a classical limit.
On the other hand, we do have 
\be
\lim_{\dimt\rightarrow 0} G_t(p,q) = H_t(p,q),
\ee

In order to have all the ingredients to build a quantum identity which resembles
formally to  a Jarzynski identity, we  need  trajectories in the Weyl space
(along which the  pseudo-work is integrated). We now explain how to construct these
 trajectories with the help of techniques developed in  \cite{RioAlm02}.

\subsection{Trajectories in  Weyl space}
\label{interacpic}
A key ingredient of  the Jarzynski identity \cite{Jar97,JarzPRE}
is the power $\der_t H_t(p(t),q(t))$ along a classical trajectory $(p(t),q(t))$, whose integral over $t$ gives the work. The exponential of the  Jarzynski  work
relates  the initial  distribution $\Pi_0(p,q)=e^{-\beta H_0(p,q)}$ to the distribution at the final time $\Pi_t(p(t),q(t))$ defined as 
$$\Pi_t(p(t),q(t))  \equiv \Pi_t(p(t),q(t) |p,q)  =e^{-\beta H_t(p(t),q(t))}$$
where  $(p(t),q(t))$ is the image under the Hamiltonian flow of the initial
point   $(p(0),q(0))=(p,q)$ in the phase space.
Now, if $\op \Pi_t = e^{-\beta\op H_t}$ is the, not normalized, quantum thermal density operator, and if the Weyl symbol of $\op \Pi_t$ is $e^{\Gamma_t(p,q)}$, from equation~(\ref{defGamma}), we would like to translate $\Pi_t(p(t),q(t))$ as a kind of propagation of $e^{\Gamma_t(p,q)}$ in phase space.

We can first remark that $\Gamma_t(p,q)$, considered as a Hamiltonian, can generate trajectories $\left(p_\Gamma(t),q_\Gamma(t)\right)$ in phase space. Then, $e^{\Gamma_t\left(p_\Gamma(t),q_\Gamma(t)\right)}$ could serve as a backbone for a Jarzynski equality in the Weyl representation, by formally reproducing the initial Jarzynski proof as if $\Gamma_t(p,q)$ was a classical Hamiltonian ( see appendix \ref{other} ). However this formal dynamics in phase space has no connection with the quantum evolution, which seems desirable if we want to obtain a quantum Jarzynski identity with physical meaning.
For this reason, in order to find a satisfactory translation of $\Pi_t(p(t),q(t))$ in the Weyl representation, we will focus on the fact that it is simply the Liouville propagation of the classical distribution $\Pi_t(p,q)$, that is
\begin{align}
\frac{d}{dt} \left(\Pi_t(p(t),q(t))\right) &= \der_t \Pi_t(p(t),q(t)) + \der_p \Pi_t(p(t),q(t))\der_t p(t) + \der_q\Pi_t(p(t),q(t))\der_t q(t) \cr
~ &= \der_t \Pi_t(p(t),q(t)) - \der_p \Pi_t(p(t),q(t))\der_q H_t(p(t),q(t)) + \der_q\Pi_t(p(t),q(t))\der_p H_t(p(t),q(t)) \cr
~ &=  \der_t \Pi_t(p(t),q(t)) - \left\{ H_t,\Pi_t\right\}(p(t),q(t)).
\label{propagLiouville}
\end{align}
This Liouville propagation will then be translated into the quantum unitary propagation $\op U_t^{\dagger} \op \Pi_t \op U_{t}$  of $\op \Pi_t$, and we shall use the fact that any quantum evolution in the Weyl representation can be described, in the semiclassical limit, as a generalized Liouville propagation, see \cite{RioAlm02}. We can find indeed a trajectory $(\check p(t),\check q(t))$ in phase space, such that
\be
\symb{\op U_t^{\dagger} \op A \op U_{t}}(p,q) \simeq \symb{\op A}\left(\check p(t),\check q(t)\right)
\ee 
where $\left(\check p(t),\check q(t)\right)$ tends to the classical trajectory $(p(t),q(t))$ when $\hbar\rightarrow 0$.
To resume, the solution $\Pi_t(p(t),q(t))$ of (\ref{propagLiouville}) will be mapped to $e^{\Gamma_t(\check p(t),\check q(t))}$, which is the approximate Weyl representation of $\op U_t^{\dagger} \op \Pi_t \op U_{t}$, solution of
\begin{align}
\frac{d}{dt} \left( \op U_t^{\dagger} \op \Pi_t \op U_{t}\right) &= \op U_t^{\dagger} \left(\der_t \op \Pi_t\right) \op U_{t} - \op U_t^{\dagger}\frac{1}{i\hbar}\left[ \op H_t,\op \Pi_t \right]\op U_{t}.
\label{quant-evol}
\end{align}
This path $(\check p(t),\check q(t))$ is
defined as in \cite{RioAlm02}, the difference being that $e^{\Gamma_t(\check p(t),\check q(t))}$ is interpreted as an imaginary time propagator. The original construction of \cite{RioAlm02} is thus shifted into the complex domain, and$(\check p(t),\check q(t))$ appears as the middle curve of two complex (nonreal) classical trajectories.

We begin  with the Weyl symbol of $\op U(\ti,\tf)^\dagger e^{-\beta \op H_{\tf}} \op U(\ti,\tf)$.
Once again we use the matrix element (\ref{VanVleck}) into the Weyl transform (\ref{weyltrans}), in real time for the real propagation, and in imaginary time for the temperature propagation,
\begin{align}
\symb{\op U(\ti,\tf)^\dagger e^{-\beta \op H_{\tf}} \op U(\ti,\tf) }(p,q) &= \int e^{-\frac{i}{\hbar}pQ}\matel {q+\frac{Q}{2}} {\op U(\ti,\tf)^\dagger e^{-\beta \op H_{\tf}} \op U(\ti,\tf)} {q-\frac{Q}{2}}~dQ \cr
&= \iiint e^{-\frac{i}{\hbar}pQ}\matel {q+\frac{Q}{2}} {\op U(\ti,\tf)^\dagger}{q_1}\matel {q_1}{e^{-\beta \op H_{\tf}}}{q_2}\matel{q_2}{\op U(\ti,\tf)} {q-\frac{Q}{2}}~dQdq_1dq_2 \cr
&= \iiint e^{-\frac{i}{\hbar}pQ}\overline{\left( \matel {q_1} {\op U(\ti,\tf)}{q+\frac{Q}{2}}\right)}\matel {q_1}{e^{-\beta \op H_{\tf}}}{q_2}\matel{q_2}{\op U(\ti,\tf)} {q-\frac{Q}{2}}~dQdq_1dq_2 \cr
~ &\simeq \iiint e^{\frac{i}{\hbar}\left\{ S_{\textrm{tot}}(Q,q_1,q_2,\tf) \right\}}~dQdq_1dq_2, 
\label{semicpropagpi}
\end{align}
where $\overline z$ is the complex conjugate of $z$, and the total action $S_{\textrm{tot}}$ is now
\be
S_{\textrm{tot}}(p,q,Q,q_1,q_2,\ti,\tf) = -pQ - S_-\left(q+\frac{Q}{2},\ti;q_1,\tf\right)
+ S_{\tf}\left(q_2,\uc_\0;q_1;\uc_\1\right) + S_+\left(q-\frac{Q}{2},\ti;q_2,\tf\right),
\label{Stot2}
\ee
with $\uc_\1=\uc_\0+\dimt \in i\mathbb R$.
Hence the total action is made of the real time action $S_+$ counted positively, the complex time action $S_{\tf}$ and the real time action $S_-$ counted negatively. We warn the reader again that the complex time action is not the analytical continuations of the real time ones, as already  mentioned in section \ref{timedep}. 

The action $S_+(q-\frac{Q}{2},\ti;q_2,\tf) = \int_{\ti}^{\tf} \left[ p_+(t)\der_t q_+(t) - H_t(p_+(t),q_+(t))\right]dt$ is calculated along a regular classical trajectory $\left(p_+(t),q_+(t)\right)$, generated by $H_t(p,q)$ with running $t$, that is,
\be
\left\{
\begin{aligned}
q_+(\ti)&=q-\frac{Q}{2} \cr
q_+(\tf)&=q_2
\end{aligned}
\right.
\qquad
\left\{
\begin{aligned}
\der_{t}p_+(t) &= -\der_qH_t\left(p_+(t),q_+(t)\right) \cr
\der_{t}q_+(t) &= \der_pH_t\left(p_+(t),q_+(t)\right);
\label{Hamilton}
\end{aligned}
\right.
\ee
the action  $S_{\tf}(q_2,\uc_\0;q_1,\uc_\1) = \int_{\uc_\0}^{\uc_\1} \left[ p_{\tf}(\uc)\der_{\uc}q_{\tf}(\uc) - H_{\tf}(p_{\tf}(\uc),q_{\tf}(\uc)) \right]d\uc.$ is calculated along the classical trajectory $\uc\mapsto\left(p_{\tf}(\uc),q_{\tf}(\uc)\right)$ generated by $H_{\tf}(p,q)$, in which the time dependence $\tf$ has been frozen.  More explicitly, we have 
\be
\left\{
\begin{aligned}
q_{\tf}(\uc_\0)&=q_2\cr
q_{\tf}(\uc_\1)&=q_1
\end{aligned}
\right.
\qquad
\left\{
\begin{aligned}
\der_{\uc}p_{\tf}(\uc) &= -\der_qH_{\tf}\left(p_{\tf}(\uc),q_{\tf}(\uc)\right) \cr
\der_{\uc}q_{\tf}(\uc) &= \der_pH_{\tf}\left(p_{\tf}(\uc),q_{\tf}(\uc)\right),
\label{imHamilton}
\end{aligned}
\right.
\ee
where $\uc$ is imaginary.
Finally, $S_-(q_1,\ti;q+\frac{Q}{2},\tf) = \int_{\ti}^{\tf} \left[ p_-(t)\der_t q_-(t) - H_{t}(p_-(t),q_-(t))\right]dt$, is calculated along a regular classical trajectory $\left(p_-(t),q_-(t)\right)$, generated by $H_{t}(p,q)$ with running $t$, that is,
\be
\left\{
\begin{aligned}
q_-(\ti) &= q+\frac{Q}{2}\cr
q_-(\tf) &= q_1
\end{aligned}
\right.
\qquad
\left\{
\begin{aligned}
\der_{t}p_-(t) &= - \der_qH_{t}\left(p_-(t),q_-(t)\right) \cr
\der_{t}q_-(t) &= \der_pH_{t}\left(p_-(t),q_-(t)\right).
\label{negHamilton}
\end{aligned}
\right.
\ee
Notice that $ - S_-\left(q+\frac{Q}{2},\ti;q_1,\tf\right)$ can be interpreted both as the opposite of an action with forward trajectory, describing the matrix element $\overline{\left( \matel {q_1} {\op U(\ti,\tf)}{q+\frac{Q}{2}}\right)}$; and as a positive action along a backward trajectory, $ - S_-\left(q+\frac{Q}{2},\ti;q_1,\tf\right) = \tilde S_-\left(q_1,\ti;q+\frac{Q}{2},\tf\right)$, describing the matrix element $\matel {q+\frac{Q}{2}} {\op U(\ti,\tf)^\dagger}{q_1}$. We found that the former way makes calculations easier to follow.

\begin{figure}[!ht]
\begin{center}
\scalebox{0.5}{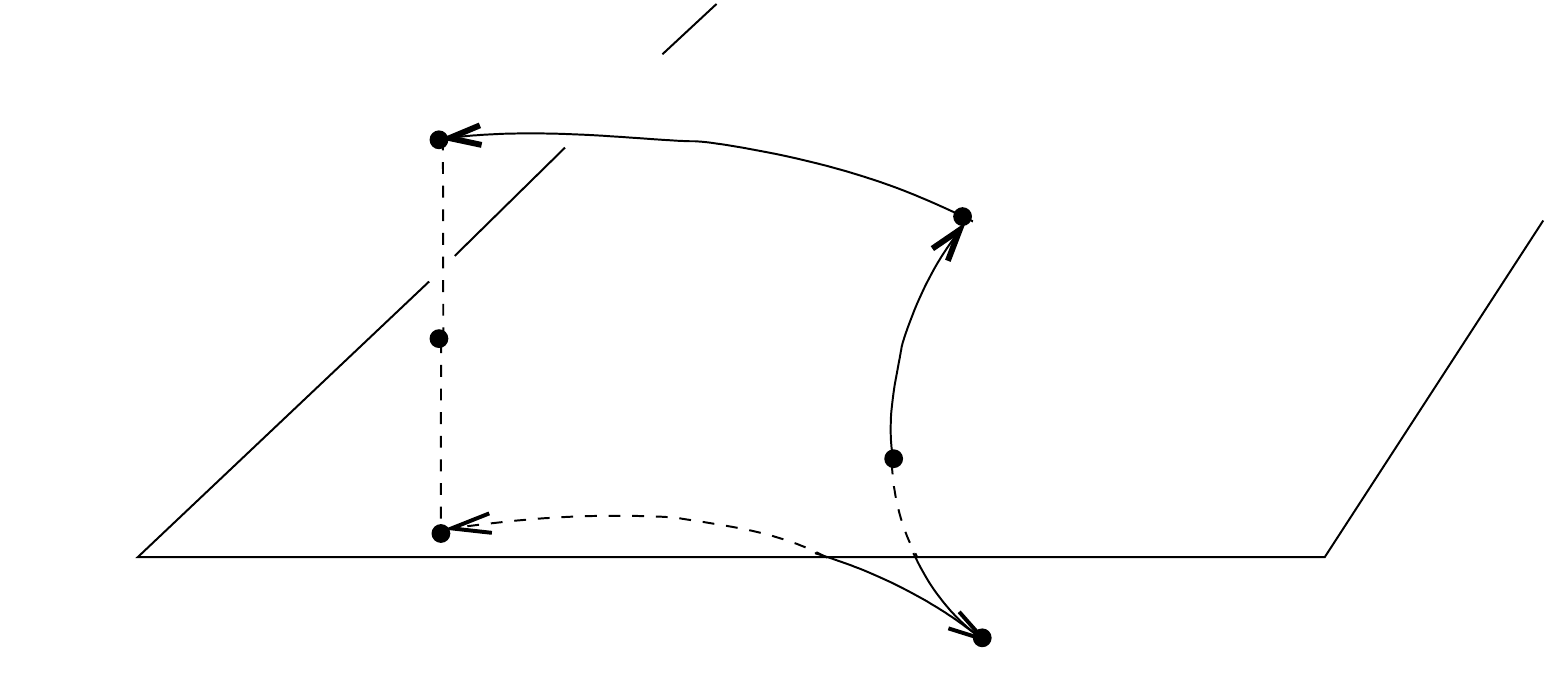}
\caption{The application $\map {\ti} {\tf} {\frac{\dimt}{2}}$ maps $(\psp_c,\qsp_c)$ to $\map{\ti} {\tf} {\frac{\dimt}{2}}(\psp_c,\qsp_c) =(p,q)$. }
\label{mapstat}
\end{center}
\end{figure}

We evaluate now expression (\ref{semicpropagpi}) with stationary phase method, that is, we replace the integral over $Q,q_1,q_2$ by a single value of the integrand at the generically unique triple $(\Qsp,\qunsp,\qdeuxsp)$ such that
\begin{align}
\frac{\der}{\der Q}S_{\textrm{tot}}(p,q,\Qsp,\qunsp,\qdeuxsp,\ti,\tf) &= -\frac{1}{2} \der_{q_\0} S_-\left(q+\frac{\Qsp}{2},\ti;\qunsp,\tf\right) - \frac{1}{2} \der_{q_\0} S_+\left(q-\frac{\Qsp}{2},\ti;\qdeuxsp,\tf\right) - p 
= 0 \label{spp2a} \\
\frac{\der}{\der q_2}S_{\textrm{tot}}(p,q,\Qsp,\qunsp,\qdeuxsp,\ti,\tf) &= \der_{q_\0} S_{\tf}\left(\qdeuxsp,\uc_\0;\qunsp;\uc_\1\right) + \der_{q_\1} S_+\left(q-\frac{\Qsp}{2},\ti;\qdeuxsp,\tf\right) = 0 
\label{spp2c} \\
\frac{\der}{\der q_1}S_{\textrm{tot}}(p,q,\Qsp,\qunsp,\qdeuxsp,\ti,\tf) &= -\der_{q_\1} S_-\left(q+\frac{\Qsp}{2},\ti;\qunsp,\tf\right) + \der_{q_\1} S_{\tf}\left(\qdeuxsp,\uc_\0; \qunsp;\uc_\1\right)  = 0 \label{spp2b}
\end{align}
This defines three trajectories, 
\begin{itemize}
\item $\left(\pspp(t),\qspp(t)\right)$, associated with $S_+\left(q-\frac{\Qsp}{2},\ti;\qdeuxsp,\tf\right)$, solution of (\ref{Hamilton}) with $\qspp(\ti)=q-\frac{\Qsp}{2}$ and $\qspp(\tf)=\qdeuxsp$,
\item $\left(\psp_{\tf}(\uc),\qsp_{\tf}(\uc)\right)$, associated with $S_{\tf}\left(\qdeuxsp,\uc_\0;\qunsp;\uc_\1\right)$, solution of (\ref{imHamilton}) with $\qsp_{\tf}(\uc_\0)=\qdeuxsp$ and $\qsp_{\tf}(\uc_\1)=\qunsp$,
\item $\left(\pspm(t),\qspm(t)\right)$, associated with $S_-\left(q+\frac{\Qsp}{2},\ti;\qunsp,\tf\right)$, solution of (\ref{negHamilton}) with $\qspm(\tf)=\qunsp$ and $\qspm(\ti) = q+\frac{\Qsp}{2}$,
\end{itemize}
such that, according to  equation (\ref{spp2a}), and using (\ref{pipf}), we have 
\begin{align}
\frac{\pspm(\ti)+\pspp(\ti) }{2} &= p \cr \hbox{ and } \,\, \quad  
\frac{\qspm(\ti)+\qspp(\ti) }{2} &= \frac{q+\frac{\Qsp}{2}+q-\frac{\Qsp}{2}}{2} = q \, . 
\label{spc1}
\end{align}
Then, from equation  (\ref{spp2c}) with (\ref{pipf}), we obtain 
\begin{align}
\pspp(\tf) &=  \psp_{\tf}(\uc_\0) \cr
\qspp(\tf) &=  \qsp_{\tf}(\uc_\0) = \qdeuxsp.
\label{connectp}
\end{align}
And  equation (\ref{spp2b}) with (\ref{pipf}) imply
\begin{align}
\pspm(\tf) &=  \psp_{\tf}(\uc_\1) \cr
\qspm(\tf) &=  \qsp_{\tf}(\uc_\1) = \qunsp.
\label{connectm}
\end{align}
To construct this set of three trajectories, one proceed as in the time independent case. We define the complex time propagator $\improp t {\uc}$ such that 
\be
\begin{aligned}
\improp {t} 0 (p',q') &= (p',q')
\end{aligned}
\qquad
\begin{aligned}
\der_\uc \improp {t} \uc (p',q') &= \mat J \nabla H_t\left(\improp {t} \uc (p',q')\right)\; ,
\end{aligned}
\ee
with $\mat J = \left( \begin{array}{cc} 0 & -1 \cr 1 & 0 \end{array} \right)$,
and the real time classical propagator $\reprop {t'} {t}$, such that 
\be
\begin{aligned}
\reprop {t'} {t'} (p',q') &= (p',q')
\end{aligned}
\qquad
\begin{aligned}
\der_t \reprop {t'} t (p',q') &= \mat J \nabla H_t\left(\reprop {t'} t (p',q')\right) \; .
\end{aligned}
\ee
We already saw in equation (\ref{conjmap}) of section \ref{sec:semiclass} that, for real $(p',q')$, the positive imaginary time propagation is the complex conjugate of the negative imaginary time propagation, that is,
\be
\improp t {-\uc} (p',q') = \overline {\improp t {\uc} (p',q') }.
\ee
On the other hand, for any complex $(p'',q'')$, one has
\be
\reprop {\ti} t (\overline {p''},\overline {q''}) = \overline {\reprop {\ti} t (p'',q'') },
\ee
since the Hamiltonian $H_t$ and times $t$ and $\ti$ are real, that is, the real time propagation conserves the complex conjugate relation between two phase space points. As a consequence, the map $\map{\ti} {\tf} {\frac{\dimt}{2}}$, defined by
\be
(p',q')\mapsto \map {\ti} {\tf} {\frac{\dimt}{2}} (p',q') = 
\frac{\left(\reprop {\ti} {\tf}\right)^{-1} \left(\improp t {\frac{\dimt}{2}} (p',q') \right)  + \left(\reprop {\ti} {\tf}\right)^{-1} \left(\improp t {-\frac{\dimt}{2}} (p',q') \right)}{2},
\ee
is a real map, that is, it maps the real phase space $\left\{(p',q')\in \mathbb R^2\right\}$ to itself. It is then easy to define the inverse image $(\psp_c,\qsp_c)$ of $(p,q)$ through map $\map {\ti} {\tf} {\frac{\dimt}{2}}$, that is
\be
(\psp_c,\qsp_c) = \left( \map {\ti} {\tf} {\frac{\dimt}{2}} \right)^{-1} (p,q),
\ee
and, from this point, to define the whole stationary phase trajectory. One has indeed,
\begin{align}
(\psp_{\tf}(\uc_\0),\qsp_{\tf}(\uc_\0)) &= \improp t {-\frac{\dimt}{2}} (\psp_c,\qsp_c) \cr
(\psp_{\tf}(\uc_\1),\qsp_{\tf}(\uc_\1)) &= \improp t {\frac{\dimt}{2}} (\psp_c,\qsp_c) \cr
(\pspp(t),\qspp(t)) &= \left( \reprop t {\tf} \right)^{-1}  (\psp_{\tf}(\uc_\0),\qsp_{\tf}(\uc_\0)) \cr
(\pspm(t),\qspm(t)) &= \left( \reprop t {\tf} \right)^{-1}  (\psp_{\tf}(\uc_\1),\qsp_{\tf}(\uc_\1)).
\end{align}
The picture is represented on figure~\ref{mapstat}.

 Notice  that, $\Qsp$, $\qunsp$, $\qdeuxsp$ and, according to the above construction, $\left(\pspp(t),\qspp(t)\right)$ and $\left(\pspm(t),\qspm(t)\right)$, are all implicit functions of times $\ti$ and $\tf$. In other words, if one changes time $\ti$ or $\tf$, the whole construction of the set of three trajectories is different. But, if one changes $\ti$ or $\tf$ in a infinitesimal way, then the trajectories will only shift infinitesimally. This will simplify further calculations because of the stationarity of the action.
\hfill\break
\noindent
{\it Remark:} although $\left(\pspm(\tf),\qspm(\tf)\right)$ is the propagation of $\left(\pspp(\tf),\qspp(\tf)\right)$ with complex time $\dimt$ and Hamiltonian $H_{\tf}$, $\left(\pspp(\ti),\qspp(\ti)\right)$ is not equal to the propagation of $\left(\pspm(\ti),\qspm(\ti)\right)$ with complex time $-\dimt$ and Hamiltonian $H_{\ti}$, because (\ref{imHamilton}) does not commute with (\ref{Hamilton}), as  explained in the paragraph before (\ref{analytHam}).
If they did  commute then the whole action would be the integral of an exact 1 form, so it would only depend on initial and final conditions $\left(\pspp(\ti),\qspp(\ti)\right)$ and $\left(\pspm(\ti),\qspm(\ti)\right)$, that is, it would be possible to deform the triple trajectory to a simple arc joining $\left(\pspp(\ti),\qspp(\ti)\right)$ to $\left(\pspm(\ti),\qspm(\ti)\right)$.  See the illustration on figure~\ref{mapstat3}. For additional remarks on this complex structure, see appendix \ref{complex}.
\begin{figure}[!ht]
\begin{center}
\scalebox{0.5}{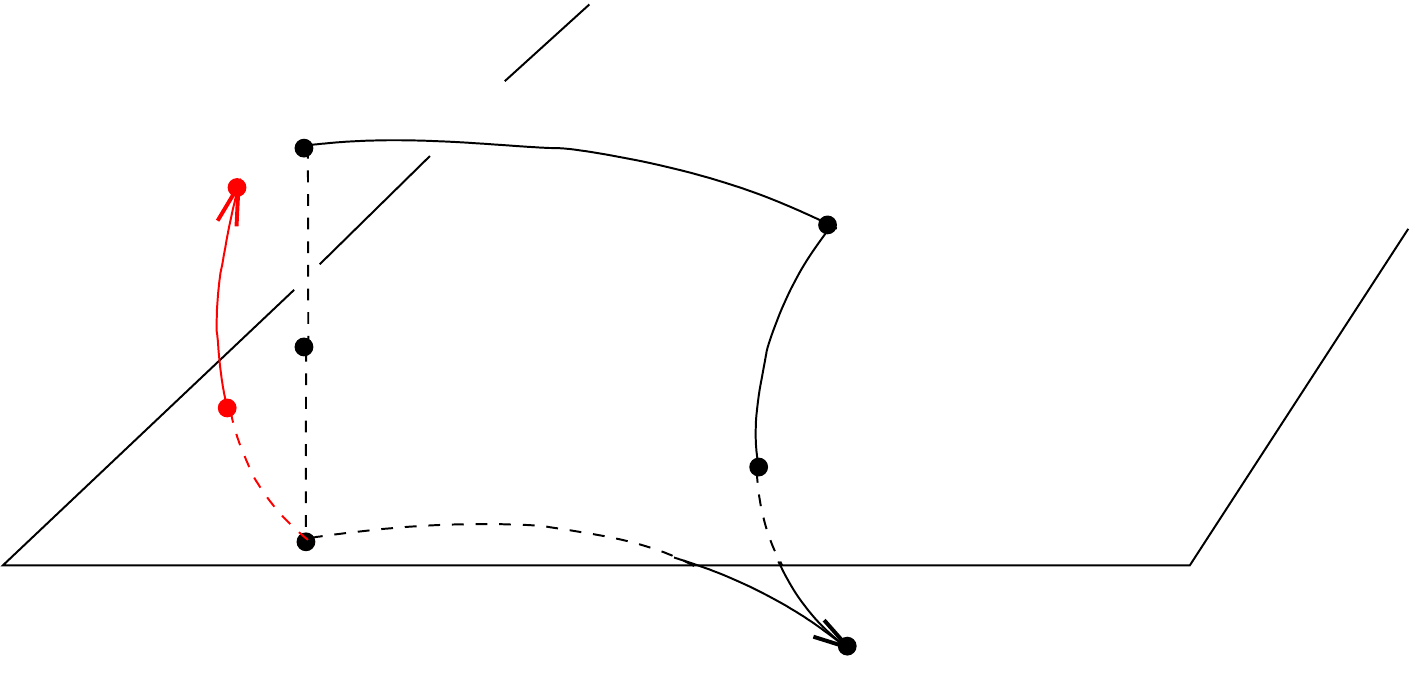}
\caption{Whereas $\left(\pspm(\tf),\qspm(\tf)\right)$ is the propagation of $\left(\pspp(\tf),\qspp(\tf)\right)$ with complex time $\dimt$, $\left(\pspp(\ti),\qspp(\ti)\right)$ is not equal to the propagation of $\left(\pspm(\ti),\qspm(\ti)\right)$ with complex time $-\dimt$. }
\label{mapstat3}
\end{center}
\end{figure}

One has finally
\begin{align}
\symb{\op U(\ti,\tf)^\dagger e^{-\beta \op H_{\tf}} \op U(\ti,\tf) }(p,q) 
~ &\simeq  e^{\frac{i}{\hbar}\left\{ -p\left(\qspp(\tf)-\qspm(\tf)\right) + \Sspp\left(\qspp(\ti),\ti;\qspp(\tf),\tf\right) 
+\Ssp_{\tf}\left(\qspp(\tf),\uc_\0;\qspm(\tf),\uc_\1\right) - \Sspm\left(\qspm(\ti),\ti;\qspm(\tf),\tf\right) \right\} }.
\end{align}
Then we define $G_{\tf}^{(\ti)}(p,q)$ by
\be
-\dimt G_{\tf}^{(\ti)}(p,q) = S_{\textrm{tot}}(p,q,\Qsp,\qunsp,\qdeuxsp,\ti,\tf).
\label{expGt1}
\ee
The superscript $(\ti)$ in $G_{\tf}^{(\ti)}(p,q)$ is meant to distinguish the argument of the exponential in $\symb{\op U(\ti,\tf)^\dagger e^{-\beta \op H_{\tf}} \op U(\ti,\tf) }(p,q)$ from the argument of the exponential in $\symb{e^{-\beta \op H_{\tf}}}(p,q)$, which would simply be $G_{\tf}(p,q)$.

\begin{figure}[ht!]
\begin{center}
\scalebox{0.5}{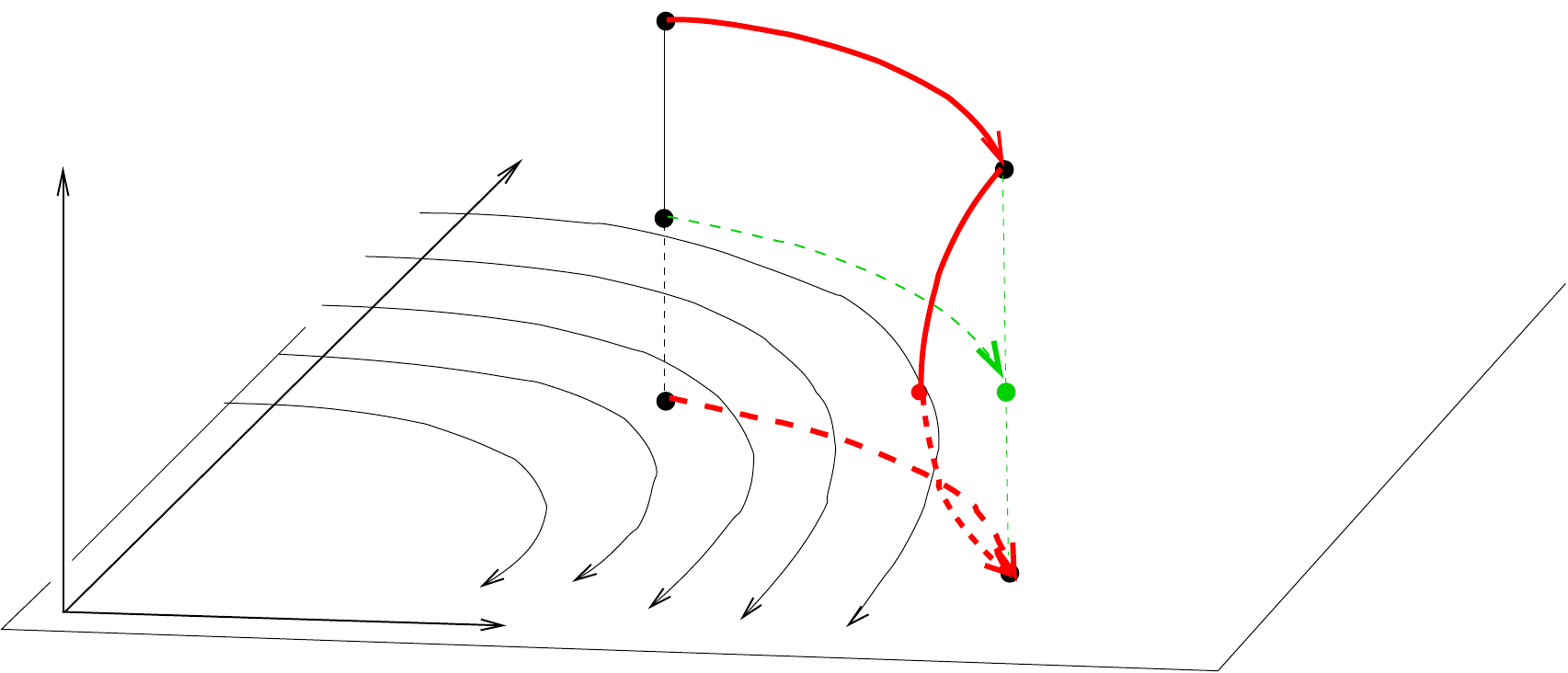}
\caption{Whereas $\der_{t}G_{t}(p,q)$ is an average over a complex arc whose middle of the chord is $(p,q)$, like in  figure~\ref{sptraj1}, $\frac{d}{d \tf} [G_{\tf}^{(\ti)}(p,q)]$ is averaged over a complex arc whose middle is the propagation of $(p,q)$ along a "pseudo classical trajectory".}
\label{trajcomp-prop}
\end{center}
\end{figure}

The time evolution of $G_{\tf}^{(\ti)}(p,q)$ can be evaluated in the same way as $G_{t}(p,q)$ in the previous subsection.  One can indeed write
\begin{align}
\frac{d}{d \tf} S_{\textrm{tot}} &= \frac{\der}{\der \tf} S_{\textrm{tot}} +  \frac{\der}{\der Q} S_{\textrm{tot}}\frac{\der \Qsp}{\der \tf} + \frac{\der}{\der q_1} S_{\textrm{tot}}\frac{\der \qunsp}{\der \tf} + \frac{\der}{\der q_2} S_{\textrm{tot}}\frac{\der \qdeuxsp}{\der \tf} \cr
~ &= \frac{\der}{\der \tf} S_{\textrm{tot}} \cr
~ &= \frac{\der}{\der \tf}\Sspp + \frac{\der}{\der \tf} \Sspm + \frac{\der}{\der \tf} \Ssp_{\tf},
\end{align}
because the $Q$, $q_1$ and $q_2$ derivatives of $S_{\textrm{tot}}$ are equal to $0$ from stationary phase conditions (\ref{spp2a}), (\ref{spp2c}) and (\ref{spp2b}). Then, from (\ref{pipf}), one has
\begin{align}
\frac{\der}{\der \tf}\Sspp\left(\qspp(\ti),\ti;\qspp(\tf),\tf\right) &= - H_{\tf}\left(\pspp(\tf),\qspp( \tf)\right) \cr
\frac{\der}{\der \tf} \Sspm\left(\qspm(\ti),\ti;\qspm(\tf),\tf\right) &=  -H_{\tf}\left(\pspm(\tf),\qspm( \tf)\right)
\end{align}
whereas, with the same arguments of stationarity used in (\ref{fluctaction}) to prove (\ref{derG}), one has
\begin{align}
\frac{\der}{\der \tf} \Ssp_{\tf}\left(\qsp_\tf(\uc_\0),\uc_\0;\qsp_\tf(\uc_\1),\uc_\1\right) &= - \int_{\uc_\0}^{\uc_\0} \der_{\tf} H_{\tf}\left(\psp_{\tf}(\uc),\qsp_{\tf}(\uc)\right)~d\tau.
\end{align}
Further, the energy is conserved along the imaginary time trajectory (\ref{imHamilton}), therefore $H_{\tf}\left(\pspp(\tf),\qspp( \tf)\right) = H_{\tf}\left(\pspm(\tf),\qspm( \tf)\right)$, keeping in mind (\ref{connectp}) and (\ref{connectm}). Therefore only remains
\begin{align}
\frac{d}{d \tf} \left( G_{\tf}^{(\ti)}(p,q)\right) &= \frac{1}{\dimt}\int_{\uc_\0}^{\uc_\1} \der_{\tf} H_{\tf}\left(\psp_{\tf}(\uc),\qsp_{\tf}(\uc)\right)~d\tau,
\label{propagpow_1}
\end{align}
We note  that it is the same expression as (\ref{derG}), except that the complex arc $\left(\psp_{\tf}(\uc),\qsp_{\tf}(\uc)\right)$ has been propagated, so $(p,q)$ is no longer the center of its chord. Let us then define $(\check p(\tf),\check q(\tf))$ as the center of the chord to the imaginary arc $\left\{ \left(\psp_{\tf}(\uc),\qsp_{\tf}(\uc)\right),\uc\in[\uc_\0,\uc_\1]\right\}$, that is, remembering (\ref{connectp}) and (\ref{connectm}),
\begin{align}
\check p(\tf) &= \frac{\psp_{\tf}(\uc_\0) + \psp_{\tf}(\uc_\1)}{2} = \frac{\pspp(\tf) + \pspm(\tf)}{2}\cr
\check q(\tf) &= \frac{\qsp_{\tf}(\uc_\0) + \qsp_{\tf}(\uc_\1)}{2}= \frac{\qspp(\tf) + \qspm(\tf)}{2}.
\label{defxcheck}
\end{align}
$(\check p(\tf),\check q(\tf))$ is then our "pseudo classical trajectory", it is like the real shadow of the couple of complex Hamiltonian trajectories $(\pspp,\qspp)$ and $(\pspm,\qspm)$, as is illustrated on figure~\ref{trajcomp-prop}. 

Hence, we have
\begin{align}
\frac{d}{d \tf} \left( G_{\tf}^{(\ti)}(p,q)\right) &=  \der_{\tf} G_{\tf}(\check p(\tf),\check q(\tf)).
\label{propagpow}
\end{align}
Equation (\ref{propagpow}) actually looks like a backward Liouville propagation
of the function defined by (\ref{derG}). The difference is that, in the Liouville case, $(p(\tf),q(\tf))$ is the classical propagation of $(p,q)$, whereas here, $(\check p(\tf),\check q(\tf))$ is the average of two classical propagations. We have
\be
\lim_{\dimt\rightarrow 0} (\check p(\tf),\check q(\tf)) = (p(\tf),q(\tf)) \,. 
\ee

\subsection{A semi-classical  Jarzynski identity}

  We  define the semi-classical work along a trajectory $(\check p(t),\check q(t))$ as
\be
\pW_{\ti,\tf}(p,q) = \int_{\ti}^{\tf} \der_{t} G_t(\check p(t),\check q(t))~dt.
\ee
According to (\ref{propagpow}) one has
\begin{align}
\pW_{\ti,\tf}(p,q) &= \int_{\ti}^{\tf} \frac{d}{dt} [G^{(\ti)}_t(p,q)]~dt \cr
~ &= G_{\tf}^{(\ti)}(p,q) - G_{\ti}(p,q)
\end{align}
that is, the pseudo-work is the difference between the final and initial pseudo-energies. This allows us to follow  the steps of the
original Jarzynski proof in \cite{Jar97}. Let $G_t(p,q)$ be the pseudo-Hamiltonian associated with $H_t(p,q)$, and
$$Z_{\textrm{sc}}(t)=\iint e^{-\beta G_t(p,q)} ~dpdq \equiv e^{-\beta F_{\textrm{sc}}(t)} \, .$$
The  average of the exponential of the pseudo-work $\pW$, in the sense defined by the expression (\ref{prodscalW}), is given by 
\begin{align}
\langle e^{-\beta \pW_{\ti,\tf}} \rangle_{\textrm{sc}} &= \iint \rho_{\tf}(p,q) ~e^{-\beta \pW(p,q)}~dp~dq \cr
~ &= \iint \frac{e^{-\beta G_{\ti}(p,q)}}{Z_{\textrm{sc}}(\ti)} ~e^{-\beta \int_{\ti}^{\tf} \der_t G_t(\check p(t),\check q(t))~dt } ~dp~dq \cr
~ &= \iint \frac{e^{-\beta G_{\ti}(p,q)}}{Z_{\textrm{sc}}(\ti)} ~e^{-\beta \left( G_{\tf}^{(\ti)}(p,q) - G_{\ti}(p,q) \right)  } ~dp~dq \cr
~ &= \iint \frac{1}{Z_{\textrm{sc}}(0)} ~e^{-\beta G_{\tf}^{(\ti)}(p,q)} ~dp~dq \cr
~ &= \frac{Z_{\textrm{sc}}(\tf)}{Z_{\textrm{sc}}(\ti)} \cr
~ &= e^{-\beta\Delta F_{\textrm{sc}}}.
\label{JarzWeyl}
\end{align}
with
\begin{align}
Z_{\textrm{sc}}(\tf) &= \iint e^{-\beta G_{\tf}^{(\ti)}(p,q)} ~dp~dq \mathop{\simeq}_{\hbar\rightarrow 0} \iint e^{-\beta \Gamma_{\tf}^{(\ti)}(p,q)} ~dp~dq = \tr{\left[\op U(\ti,\tf)^\dagger e^{-\beta \op H_{\tf}}\op U(\ti,\tf)\right]} = \tr{\left[e^{-\beta \op H_{\tf}}\right]} \cr
Z_{\textrm{sc}}(\ti) &= \iint  e^{-\beta G_{\ti}(p,q)} ~dp~dq \mathop{\simeq}_{\hbar\rightarrow 0} \iint e^{-\beta \Gamma_{\ti}(p,q)} ~dp~dq =  \tr{\left[e^{-\beta \op H_{\ti}}\right]}.
\end{align}

Finally, we have  obtained a  Jarzynski identity in the semi-classical limit
\be
\boxed{
\iint \frac{e^{-\beta G_{\ti}(p,q)}}{Z_{\textrm{sc}}(\ti)} e^{-\beta \pW_{\ti,\tf}(p,q) } ~dp~dq = \frac{Z_{\textrm{sc}}(\tf)}{Z_{\textrm{sc}}(\ti)}
}
\label{JarzWeylF}
\ee
where the pseudo-work $\pW_{\ti,\tf}(p,q) = \int_{\ti}^{\tf} \der_{t} G_t(\check p(t),\check q(t))~dt$ is evaluated along the pseudo-trajectory $(\check p(t),\check q(t))$ which starts at $(p,q)$, with a probability given by the thermal state $\frac{e^{-\beta G_{\ti}(p,q)}}{Z(\ti)}$. The average of the exponential of this pseudo-work then gives the ratio of the quantum partition functions in the semi-classical limit.

\section{A solvable example:  the harmonic oscillator}
\label{harmosc}

It is useful to illustrate  formal non-equilibrium  identities on 
some specific systems. In the case of
the quantum work relations, exactly  solvable models  are rare and  mostly limited to the harmonic
oscillator,   non-interacting quantum gases, or  two-level systems
  \cite{JarzQuan,Liu14,Mahl07,Lutz08b,Hangg08,Plast13}. 
  Here, we shall  apply  the formulas of  the previous sections to the
  harmonic oscillator,  whose classical Hamiltonian is defined by
\be
H(p,q)=\frac{1}{2m}p^2+\frac{1}{2}m\omega^2 q^2 \, 
\label{oscharm}
\ee
and the  corresponding quantum Hamiltonian is
\be
\op H = -\frac{\hbar^2}{2m}\der_q^2 + \frac{1}{2}m\omega^2 q^2.
\ee

\subsection{Exact canonical distribution in Weyl representation}

In order to calculate the  canonical distribution in Weyl representation, we 
decompose the density operator of the thermal state in the Fock states $|n\rangle$, and then
use the Weyl representation of a projector on a Fock state, $\ket n \bra n$, which is known to be
\cite{Curt,Alm98}:
\begin{align}
W_n(p,q) &=  \frac{(-1)^n}{\pi\hbar}e^{-\frac{2}{\hbar\omega}H(p,q)}L_n\left(\frac{4}{\hbar\omega}H(p,q)\right).
\end{align}
Then, we have 
\be
e^{-\beta \op H} = \sum_n e^{-\beta \hbar \omega(n+1/2)} |n\rangle\langle n| .
\label{pi}
\ee
It  is also useful to introduce
\be
e^{\beta \op H} = \sum_n e^{\beta \hbar \omega(n+1/2)} |n\rangle\langle n|.
\label{pi-1}
\ee

~

By virtue of the linearity of the Weyl transform, the Weyl symbol of $e^{-\beta \op H}$ is
given by 
\begin{align}
\left[e^{-\beta \op H}\right]_W(p,q) &= \sum_n e^{-\beta \hbar \omega(n+1/2)} W_n(p,q) \cr
~ &= \sum_n e^{-\beta \hbar \omega(n+1/2)}\frac{(-1)^n}{\pi \hbar}e^{-\frac{2}{\hbar\omega}H(p,q)}L_n\left(\frac{4}{\hbar\omega}H(p,q)\right),
\end{align}
where we recognize the generating function of the Legendre polynomials,
\be
\sum_n X^n L_n(Y) = \frac{e^{-\frac{YX}{1-X}}}{1-X},
\label{generat}
\ee
from which we deduce that 
\be
\boxed{
\symb{e^{-\beta \op H}}(p,q)  = \frac{1}{2\pi\hbar\cosh{\left(\frac{\beta\hbar\omega}{2}\right)}}
\exp{\left[-\frac{2}{\hbar\omega} \tanh{\left(\frac{\beta\hbar\omega}{2}\right)} H(p,q) \right]}.
}
\label{piW}
\ee

\subsection{Canonical distribution from the stationary phase approximation}

Dealing with a harmonic oscillator, we expect the stationary phase method to be exact.
The imaginary time trajectory can simply be obtained. Starting from the real intersection of the complex arc with the real plane, $(\psp_c,\qsp_c)$, it is given by 
\begin{align}
\psp(\uc) &= \psp_c \cos{\omega \uc} - m\omega \qsp_c \sin{\omega \uc} \cr
\qsp(\uc) &= \qsp_c \cos{\omega \uc} + \frac{\psp_c}{m\omega}\sin{\omega \uc},
\end{align}
Hence, from (\ref{thermaction}), where the above choice of origin implies that $\uc_\0 = -\frac{\dimt}{2}$ and $\uc_\1 = \frac{\dimt}{2}$, we obtain 
\begin{align}
S &= -i\left( \frac{(\psp_c)^2}{2m} -\frac{1}{2}m\omega^2 (\qsp_c)^2 \right) \frac{\sinh{\omega \hbar\beta}}{\omega} \cr
~ &= \frac{im\omega}{2}\frac{\cosh{(\hbar\beta\omega)}((\qsp_\1)^2+(\qsp_\0)^2) - 2\qsp_\1\qsp_\0}{\sinh{(\hbar\beta\omega)}}
\end{align}
On the other hand, from the relation 
\begin{align}
p\Qsp&= \frac{\psp(\frac{\dimt}{2}) + \psp(-\frac{\dimt}{2})}{2}\left(\qsp\left(\frac{\dimt}{2}\right) - \qsp\left(-\frac{\dimt}{2}\right)\right), 
\end{align}
we deduce 
\be
\frac{i}{\hbar}\left(\Ssp - p\Qsp\right) = -\frac{1}{\hbar\omega} \sinh{(\omega \hbar\beta)} \left( \frac{(\psp_c)^2}{2m} + \frac{1}{2}m\omega^2 (\qsp_c)^2\right).
\ee
Noting  that 
\begin{align}
p&=\frac{\psp(-\frac{\dimt}{2})+\psp(\frac{\dimt}{2})}{2}=\psp_c\cosh{\frac{\omega\hbar\beta}{2}}\cr
q&=\frac{\qsp(-\frac{\dimt}{2})+\qsp(\frac{\dimt}{2})}{2}=\qsp_c\cosh{\frac{\omega\hbar\beta}{2}},
\label{xx0}
\end{align}
we conclude, from (\ref{expG}) and (\ref{pref}), that 
\be
\boxed{
e^{-\beta G(p,q)} = \frac{1}{2\pi\hbar\cosh{\left(\frac{\omega\hbar\beta}{2}\right)}}\exp{\left[- \frac{2}{\hbar\omega}\tanh{\left(\frac{\beta\hbar\omega}{2}\right) } H(p,q)\right]}
}
\ee
and we  recover  expression (\ref{piW}).

\subsection{Quantum power and pseudo-work}

The Jarzynski power of the Hamiltonian (\ref{oscharm}) with time dependent $\omega$ is
given by 
\be
\der_t H_t = m\dot \omega_t \omega_t q^2.
\ee
Applying equation (\ref{derG}), we obtain 
\begin{align}
\der_tG_t(p,q) &= \frac{1}{\dimt}m\omega_t\dot\omega_t\int_{-\frac{\dimt}{2}}^{\frac{\dimt}{2}}  \left(\qsp(\uc)\right)^2~d\tau \cr
~ &= -\frac{1}{\hbar\beta}m\omega_t\dot\omega_t \int_{\frac{\hbar\beta}{2}}^{-\frac{\hbar\beta}{2}} \left[
 (\qsp_c)^2 \cosh{(\omega_t t')}^2 
- \frac{(\psp_c)^2}{m^2\omega_t^2}\sinh{(\omega_t t')}^2
+  \frac{2i}{m\omega_t}\psp_c\qsp_c\cosh{(\omega_t t')}\sinh{(\omega_t t')} \right]~ dt' \cr
~ &= \frac{1}{\hbar\beta}m\omega_t\dot\omega_t \left[
 (\qsp_c)^2 \left( \frac{\sinh{(\omega_t \hbar\beta)}}{2\omega_t} + \frac{\hbar\beta}{2} \right)
- \frac{(\psp_c)^2}{m^2\omega_t^2}\left( \frac{\sinh{(\omega_t \hbar\beta)}}{2\omega} - \frac{\hbar\beta}{2} \right)
 \right], 
\end{align}
which, using equation  (\ref{xx0}), leads to 
\be
\boxed{
\der_tG_t(p,q)  = \frac{\dot\omega_t}{\omega_t} \frac{2}{1+\cosh{\beta\hbar\omega_t}}\left[
 \frac{1}{2}m\omega_t^2 q^2 \left( \frac{\sinh{(\beta\hbar\omega_t)}}{(\beta\hbar\omega_t)} + 1 \right)
+ \frac{p^2}{2m}\left( 1 - \frac{\sinh{(\beta\hbar\omega_t)}}{\beta\hbar\omega_t} \right)
 \right].
}
\label{Pt}
\ee
One can verify that
\be
\lim_{\beta\hbar\rightarrow 0} \der_tG_t(p,q) = m\omega_t\dot\omega_t \left(\lim_{\beta\hbar\rightarrow 0} \qsp_c\right)^2 = m\omega_t\dot\omega_t q^2 = \der_t H_t(p,q)
\ee
\noindent
Remark:  we note that the pseudo-trajectory $(\check p(t),\check q(t))$ actually coincides with the classical trajectory $(p(t),q(t))$ because of   the linearity of the dynamics
of the harmonic oscillator with
respect to the initial conditions,
\be
\left(\begin{array}{c}p(t) \cr q(t) \end{array}\right)  = \left(\begin{array}{cc}a(t) & b(t) \cr c(t) & d(t) \end{array}\right) \left(\begin{array}{c} p(0) \cr q(0) \end{array}\right) .
\ee

\section{Conclusion}

We have derived a formal Jarzynski identity in  the Weyl representation, whose classical  limit is the  celebrated  nonequilibrium work  identity for  the corresponding classical Hamiltonian. This formal  identity involves the average of the exponential of a pseudo-work $\der_t G_t$ along pseudo-trajectories which start from initial phase space points distributed according to a thermal distribution. This average is  shown to be equal to the ratio of the partition functions of the semi-classical thermal distribution which is
approximately equal to the ratio of the quantum partition functions in the semi-classical regime --  as long as the semi-classical Van Vleck propagator can be considered to be valid approximation. In other words,  we have
obtained  a semi-classical estimate of the quantum free energy from a $\hbar$ correction of the Jarzynski identity which involve only classical quantities.
This relation has been verified for  the quantum harmonic oscillator.

The logical path of the present work can be summarized by the following
diagram, which  explains  that we have combined the Weyl quantization
scheme with a semi-classical (i.e. stationary phase) calculation
and draws a parallel between the   classical  derivation
of Jarzynski and its mirror image in the Weyl  quantum  phase space:
\be
\begin{CD}
e^{-\beta H_t(p,q)}    @>\textrm{Work}>> W_t(p,q) = \int_0^t \der_t H_{t'}(p(t'),q(t'))~dt' @>\textrm{Jarzynski}>> \iint \frac{e^{-\beta H_0(p,q)} }{Z_{\textrm{cl}}(0)} e^{-\beta W_t(p,q)}~dpdq = \frac{Z_{\textrm{cl}}(t)}{Z_{\textrm{cl}}(0)}      \\
@VV\textrm{Quant.}V   \\
  e^{-\beta \op H_t}  \\ 
@VV\textrm{Weyl}V   \\
e^{-\beta \Gamma_t(p,q)}  \\
@VV\textrm{semiclassical}V    \\
e^{-\beta G_t(p,q)} @>\textrm{Work}>> \pW_t(p,q) = \int_0^t \der_t G_{t'}(\check p(t'),\check q(t'))~dt' @>\textrm{Jarzynski}>> \iint \frac{e^{-\beta G_0(p,q)} }{Z_{\textrm{sc}}(0)} e^{-\beta\pW_t(p,q)}~dpdq = \frac{Z_{\textrm{sc}}(t)}{Z_{\textrm{sc}}(0)}
\end{CD}
\ee

A natural extension of this study would be to use the formal  identity
in the  Weyl space  to calculate quantum corrections to the classical Jarzynski
formula --  order by order with respect to $\hbar$ --  and to find
 some geometric interpretation of these corrections. It would  also be
 interesting to compare the pseudo-work defined here 
with the work operator defined for Lindblad equations \cite{CheMal}, in 
the markovian framework of an  open system constantly monitored by its environment.

\appendix
\section{Other trajectories}
\label{other}
The approach we followed led us to define a pseudo-Hamiltonian $\Gamma_t(p,q)$, whose semiclassical approximation is $G_t(p,q)$. It might have seemed natural to use this pseudo-Hamiltonian as a proper classical-like Hamiltonian, and generate real classical trajectories $(p_r(t),q_r(t))$ directly from Hamilton equations
\begin{align}
\dot p_r(t) &= -\frac{\der \Gamma_t}{\der q}(p_r(t),q_r(t)) \cr
\dot q_r(t) &= \frac{\der \Gamma_t}{\der p}(p_r(t),q_r(t)).
\end{align}
From the structure of these equations, we would then automatically have
\be
\frac{d}{dt} \Gamma_t(p_r(t),q_r(t)) = \frac{\der}{\der t} \Gamma_t(p_r(t),q_r(t)),
\ee
which is enough to derive a Jarzynski identity, by following exactly the classical proof for a Hamiltonian closed system as presented in \cite{Jar97}. It would give simply, by using the same argument of symplecticity of trajectories,
\begin{align}
\int \frac{e^{-\beta\Gamma_0(p_r(\ti),q_r(\ti))}}{Z(\ti)}e^{-\beta\int_{\ti}^{\tf}\frac{\der}{\der t} \Gamma_t(p_r(t),q_r(t))~dt}~dp_r(\ti)dq_r(\ti) &= \frac{1}{Z(\ti)}\int e^{-\beta\Gamma_0(p_r(\tf),q_r(\tf))} ~dp_r(\ti)dq_r(\ti) \cr
~ &= \frac{1}{Z(\ti)}\int e^{-\beta\Gamma_0(p_r(\tf),q_r(\tf))} ~dp_r(\tf)dq_r(\tf) \cr
~ &= \frac{Z(\tf)}{Z(\ti)},
\end{align}
where the second line uses the fact that the Jacobian of $(p_r(\ti),q_r(\ti))\mapsto (p_r(\tf),q_r(\tf))$ is equal to $1$, because of symplecticity.
 
We didn't retain this option because, although the proof is formally correct, we could not find any quantum signification to these trajectories in terms of the Weyl representation of a quantum evolution. On the other hand, our pseudo-trajectory $(\check p(t),\check q(t))$ naturally describes the Weyl representation of the quantum evolution of an operator, in the semiclassical limit. We have indeed,
\be
\symb{\op U(\ti,\tf)^\dagger \op A \op U(\ti,\tf) }(p,q) = A(\check p(\tf),\check q(\tf)). 
\ee

The example of the harmonic oscillator $H_t(p,q)=\frac{p^2}{2m}+\frac{1}{2}m\omega_t^2 q^2$ is quite suggestive here. The pseudo-Hamiltonian $\Gamma_t(p,q)$ is just the proper Hamiltonian multiplied by $\lambda =\frac{1}{\hbar\omega}\tanh{\frac{\beta\hbar\omega}{2}}$, and therefore the classical-like trajectory generated by $\Gamma_t(p,q)$ is a classical trajectory of the Harmonic oscillator, but where the time has been scaled by $\lambda$. This obviously comes from the Hamilton equations,
\begin{align}
\dot p_r(t) &= -\lambda \frac{\der H_t}{\der q}(p_r(t),q_r(t)) \cr
\dot q_r(t) &= \lambda\frac{\der H_t}{\der p}(p_r(t),q_r(t)),
\end{align}
which can be written
\begin{align}
\frac{\der{p_r}}{\der (\lambda t)} &= - \frac{\der H_t}{\der q}(p_r(\lambda t),q_r(\lambda t)) \cr
\frac{\der{q_r}}{\der (\lambda t)} &= \frac{\der H_t}{\der p}(p_r(\lambda t),q_r(\lambda t)).
\end{align}
 
On the other hand, the pseudo-trajectory $(\check p(t),\check q(t))$ is exactly the classical trajectory $(p(t),q(t))$ with the correct time $t$, 
\begin{align}
\dot p(t) &= - \frac{\der H_t}{\der q}(p(t),q(t)) \cr
\dot q(t) &= \frac{\der H_t}{\der p}(p(t),q(t)),
\end{align}
which suggests that this choice is indeed more natural.

\section{Complex symplectic structure}
\label{complex}
The complex structure of the trajectories is made of two types of trajectories, the imaginary 'formal time' trajectories $\left(\psp_{\tf}(\uc),\qsp_{\tf}(\uc)\right)$, and the real time ones, $\left(\pspp(t),\qspp(t)\right)$ and $\left(\pspm(t),\qspm(t)\right)$. They do not commute, therefore the mixed complex phase space trajectories that we consider cannot be considered as the analytical continuation of classical phase space $(p(t),q(t))$. 

Obviously, freezing the time real $t$ of the Hamiltonian $H_t(p,q)$ and propagating through Hamiltonian dynamics with complex formal time, like for instance (\ref{imHamilton}) if one restricts to purely imaginary formal time, will generate trajectories whose reunion is the manifold $H_t(p,q)=E$, where $E$ is a complex constant. For instance, in the scheme of FIG. 4, the energy $E_c=H_{\ti}\left(\psp_{\tf}(\uc),\qsp_{\tf}(\uc)\right)$, along the arc which goes from $\left(\psp_{\tf}(\uc_\0),\qsp_{\tf}(\uc_\0)\right)$ to $\left(\psp_{\tf}(\uc_\1),\qsp_{\tf}(\uc_\1)\right)$, is a real constant, because the arc crosses the real plane $(\mathop{Re}{(p)},\mathop{Im}{(q)})$ at $(p_c^\star,q_c^\star)$, and the Hamiltonian is real.

On the other hand, the energy is not conserved any more through the real time propagations $\left(\pspp(t),\qspp(t)\right)$ and $\left(\pspm(t),\qspm(t)\right)$, driven by (\ref{Hamilton}). Still, if we define $E_+(t)=H_t\left(\pspp(t),\qspp(t)\right)$ and $E_-(t)=H_t\left(\pspm(t),\qspm(t)\right)$, then we have $E_+(t)=\overline{E_-(t)}$, that is, both energies are complex conjugate. Also, we have $E_+(\ti)=E_-(\ti)=E_c\in \mathbb R$, that is, the energy is initially real at $t=\ti$, then they draw two complex conjugate branches until $t=\tf$. The midpoint of the final tips of these two complex conjugate branches is the real phase space point $(p,q)$ where the Weyl function is evaluated (see FIG. 4).

Let us follow for instance $\left(\pspp(t),\qspp(t)\right)$, and then let us freeze the time $t$ at a later value $t=t_1>\ti$. Then, propagating from $\left(\pspp(t_1),\qspp(t_1)\right)$ with complex formal time would generate a new set of trajectories whose reunion is the manifold $\left\{(p,q) \in{\mathbb C}^2 \textrm{~such~that~} H_{t_1}(p,q)=H_{t_1}\left(\pspp(t_1),\qspp(t_1)\right)=E_+(t_1)\right\}$, where $E_+(t_1)\in\mathbb C$. Since $H_{t_1}(p,q)\neq H_{\ti}(p,q)$, nothing prevents this new manifold to intersect the initial $E_c$ manifold.

To resume, for a fixed time $t$, the whole two dimensional complex phase space $(p,q)\in{\mathbb C}^2$ can be organized as a foliation $\mathcal F_t$ of one complex dimensional manifolds. Each one of this complex manifold is stable through the Hamiltonian dynamics generated by $H_t$ with frozen time dependence $t$, and it can be identified with $\left\{ (p,q)\in{\mathbb C}^2 \textrm{~such~that~} H_t(p,q)=E\right\}$, with some $E\in \mathbb C$, so we may call it $\mathcal L_t(E)$. The reunion of all the $\mathcal L_t(E)$ for all values of $E$ gives back $\mathbb C^2$. In particular, every imaginary formal time dynamics (\ref{imHamilton}) lives on such a manifold $\mathcal L_t(E)$. However, changing time $t$ changes the whole foliation. As a consequence, a real time trajectory $(p(t),q(t))$, defined by (\ref{Hamilton}), which crosses a leave $\mathcal L_t(E)$ at time $t$, will then cross an other leave $\mathcal L_{t+dt}(E+dE)$ at time $t+dt$, with $dE$ defined by
\begin{align}
dE &= \frac{d}{dt}H_{t}(p(t),q(t))~dt \cr
~ &= \frac{\der}{\der t}H_{t}(p(t),q(t))~dt,
\end{align}
so $\frac{dE}{dt}$ is a function of the trajectory $(p(t),q(t))$. Although $\mathcal L_{t}(E+dE)$ does not intersect $\mathcal L_{t}(E)$, $\mathcal L_{t+dt}(E+dE)$ can a priori intersect $\mathcal L_{t}(E)$ at points $\left(p_a,q_a\right)$ solutions of
\be
\begin{cases}
H_{t+dt}\left(p_a,q_a\right) = E+dE \cr
H_t \left(p_a,q_a\right) = E,
\end{cases}
\ee
that is
\be
\begin{cases}
H_t \left(p_a,q_a\right) = E \cr
\frac{\der }{\der t}H_t\left(p_a,q_a\right) = \frac{dE}{dt} = \frac{\der }{\der t}H_t(p(t),q(t)).
\end{cases}
\ee
For instance, if $H_t(\vct x)=H_0(\vct x)+\vct F_t \cdot \vct x$, where $\vct x = (p,q)$, then these equations become
\be
\begin{cases}
H_t \left(\vct x_a\right) = E \cr
\dot{\vct F}_t \cdot \vct x_a = \dot{\vct F}_t \cdot \vct x(t),
\end{cases}
\ee
which implies that $\vct x_a$ is at the intersection between the complex line $\vct x(t) + a \mat J \dot{\vct F}_t, a\in\mathbb C$ and the surface of constant energy $E$. In other words, $\vct x_a = \vct x(t) + a \mat J \dot{\vct F}_t$, where $a\in \mathbb C$ is such that
\be
H_t \left(\vct x(t) + a \mat J \dot{\vct F}_t\right) = E.
\ee
As soon as $H_t$ is quadratic or more, there can be other solutions than $a=0$.

A natural symplectic structure can be attached to each manifold $\mathcal L_{t}(E)$ by using Hamilton's equations (\ref{imHamilton}), generalized to arbitrary complex $\tau$, as Schwartz identities. On the other hand, the real time dynamics, driven by (\ref{Hamilton}), can be supplied with an other symplectic structure, adding $(E,t)$ as an extra couple of canonical variables to $(p,q)$, and this structure can also be extended to complex $(E,t)$.
However, how to connect both symplectic structures is not clear to us by now.

\acknowledgements{The authors would like to thank D.~Ullmo, S.~Tomsovic and R.~Sopracase for useful discussions and comments. We also thank the referees for very stimulating remarks.}

\end{document}